\documentclass[letter,twocolumn,10pt]{extarticle}

\usepackage[number,note]{achemso}
\usepackage{jpc}
\usepackage[super,sort&compress,comma]{natbib}

\usepackage{rotating}
\usepackage{graphicx} 
\usepackage{amssymb}
\usepackage{amsmath}
\usepackage{color}
\usepackage{comment}
\usepackage[normalem]{ulem}
\usepackage{wasysym}
\usepackage{bm}
\usepackage[pdftex,colorlinks,citecolor=blue]{hyperref}
\usepackage{dblfloatfix}

\usepackage{xr}
\newlength{\swid}
\newlength{\dwid}
\renewcommand{\theenumi}{\roman{enumi}}

\definecolor{Magenta}{rgb}{0.80,0.00,0.80}
\definecolor{Blue}{rgb}{0.00,0.00,0.90}
\definecolor{Green}{rgb}{0.00,0.55,0.00}
\definecolor{Red}{rgb}{0.80,0.00,0.00}
\definecolor{Brown}{rgb}{0.50,0.40,0.00}
\definecolor{Cyan}{rgb}{0.00,0.50,0.50}

\usepackage{soul}
\newcommand{\DefineAuthor}[2]{%
  \expandafter\newcommand\csname #1\endcsname[1]{\textcolor{#2}{##1}}%
  \expandafter\newcommand\csname #1note\endcsname[1]{\textcolor{#2}{\hl{\textbf{\{#1:} \textit{##1}\textbf{\}}}}}%
  \expandafter\newcommand\csname #1del\endcsname[1]{\textcolor{#2}{\sout{##1}}}%
  \expandafter\newcommand\csname #1change\endcsname[2]{\textcolor{red}{\sout{##1}} \textcolor{#2}{##2}}%
}
\DefineAuthor{tk}{Blue}    
\DefineAuthor{ma}{Magenta} 

\def\Fig{Figure}
\def\Figs{Figures}

\def\Tab{Table}
\def\Tabs{Tables}
\def\Eq{eq}

\def\obo{\ensuremath{(1\times1)}}
\def\tbt{\ensuremath{(2\times2)}}

\def\Aex{\ensuremath{A_{\mathrm{usa}}}}
\def\Aa{\ensuremath{A_{\mathrm{a}}}}
\def\Ra{\ensuremath{R_{\mathrm{a}}}}
\def\Rc{\ensuremath{R_{\mathrm{c}}}}

\def\Eb{\ensuremath{E_{\rm b}}}

\def\Eobo{\ensuremath{E^{\obo}_{\mathrm{b}}}}
\def\Etbt{\ensuremath{E^{\tbt}_{\mathrm{b}}}}
\def\Del{\ensuremath{\Delta_{\Eb}}}
\def\RQ{\ensuremath{f_{\rm rec}}}

\begin{document}

\title{\textbf{Surprising lateral interactions between negatively charged adatoms on metal surfaces}}

\author{Matic Pober\v znik and Anton Kokalj*\\[1em]
  \normalsize
  Department of Physical and Organic Chemistry,
  Jo\v zef Stefan Institute, Jamova 39, SI-1000 Ljubljana, Slovenia
}

\date{\small \today}

\twocolumn[
\maketitle
\begin{abstract}
  \begin{quote}
    Even something as conceptually simple as adsorption of
    electronegative adatoms on metal surfaces, where repulsive lateral
    interactions are expected for obvious reasons, can lead to
    unanticipated behavior. In this context, we
    explain the origin of surprising lateral interactions between
    electronegative adatoms observed on some metal surfaces by means
    of density functional theory calculations of four electronegative
    atoms (N, O, F, Cl) on 70 surfaces of 44 pristine metals. Four
    different scenarios for lateral interactions are identified, some
    of them being unexpected: (i) they are repulsive, which is the
    typical case and occurs on almost all transition metals. (ii,iii)
    They are atypical, being either attractive or negligible, which
    occurs on p-block metals and Mg, and (iv) surface reconstruction
    stabilizes the low-coverage configuration, preventing atypical
    lateral interactions. The last case occurs predominantly on
    s-block metals.


  \end{quote}
\end{abstract}
\vspace*{2em}] 

\thanks{\noindent
  \rule{0.2\textwidth}{1pt}\\
  \footnotesize
  ORCID IDs: 0000-0002-4866-4346 (MP), 0000-0001-7237-0041 (AK)\\
  * Corresponding Author: Anton Kokalj, Tel: +386-1-477-35-23; 
  Fax: +386 1 251 93 85, E-mail: tone.kokalj@ijs.si,
  URL: \texttt{http://www.ijs.si/ijsw/K3-en/Kokalj}\\
}


\section*{INTRODUCTION}

The adsorption of electronegative atoms on metal surfaces is of
paramount importance in surface science as well as
electrochemistry.\cite{Magnussen_CR102,Tripkovic_FD140,Andryushechkin_SSR73,Zhu_JESC163}
As an electronegative atom approaches a metal surface, charge is
transferred and it becomes negatively charged. This interaction can be
described classically by the method of images, where the
adatom/image-charge pair can be seen as a dipole. As more adatoms
accumulate on the surface repulsive interactions are expected between
them. Such interactions were confirmed for a variety of adatoms on
metal surfaces
\cite{Miller_JCP134,Loffreda_JCP108,Gava_PRB78,Ma_SS619,Peljhan_JPCC113,Inderwildi_JCP122,Gossenberger_SS631}
and they typically scale as
$\mu^2/R^3 \propto \mu^2\Theta^{\frac{3}{2}}$, where $\mu$ is the
adatom induced dipole, $R$ is the nearest-neighbor interadatom
distance, and $\Theta$ is the surface coverage. However, in a few
cases, notably for electronegative atoms on Mg(001)
\cite{Francis_PRB87,Cheng_PRL113} and O on
Al(111),\cite{Jacobsen_PRB52,Kiejna_PRB63,Poberznik_JPCC120}
counterintuitive attractive interactions were identified. In our
previous publication \cite{Poberznik_JPCC120} we explained that these
surprising attractive lateral interactions are a consequence of the
interplay between electrostatic and geometric effects and that there
exists a critical height of adatoms above the surface, below which
attractive interactions can emerge. Since this model---explained in
the Supporting Information and henceforth referred to as the simple
ionic model---requires only (i) sufficiently ionic bonding and (ii) a
low height of the adatom above the surface, it stands to reason that
it should be generally applicable, provided that the two requirements
are met.
To address this proposition, the adsorption of four different
electronegative adatoms (N, O, F, and Cl) on 44 elemental metals, as
indicated in \Fig~\ref{fig:ChosenSystemsPeriodicTable}, is considered
herein by means of density-functional-theory (DFT) calculations.

\section*{TECHNICAL DETAILS}
DFT calculations were performed with the {\tt PWscf} code from the
{\tt Quantum ESPRESSO} distribution\cite{Giannozzi_JPCM29} and the
{\tt PWTK} scripting environment,\cite{PWTK} using the generalized
gradient approximation (GGA) of Perdew--Burke--Ernzerhof (PBE)
.\cite{Perdew_PRL77} We used the projector augmented wave (PAW)
method\cite{Blochl_PRB50} with PAW potentials obtained from a
pseudopotential library.\cite{DalCorso_CMS95,pseudos} Kohn--Sham
orbitals were expanded in a plane wave basis set with a kinetic energy
cutoff of 50 Ry (600 Ry for the charge density).  Brillouin zone (BZ)
integrations were performed with the special point
technique,\cite{MonkhorstPack_PRB13} using a $12\times12\times1$
shifted $k$-mesh for \obo\ surface cells (or equivalent for larger
cells) and a Methfessel-Paxton smearing\cite{Methfessel_PRB40} of
0.02~Ry. Molecular graphics were produced by the {\tt XCrySDen}
graphical package.\cite{Kokalj_JMGM17}

\begin{figure*}[t]
   \centering
   \includegraphics[width=1.0\textwidth]{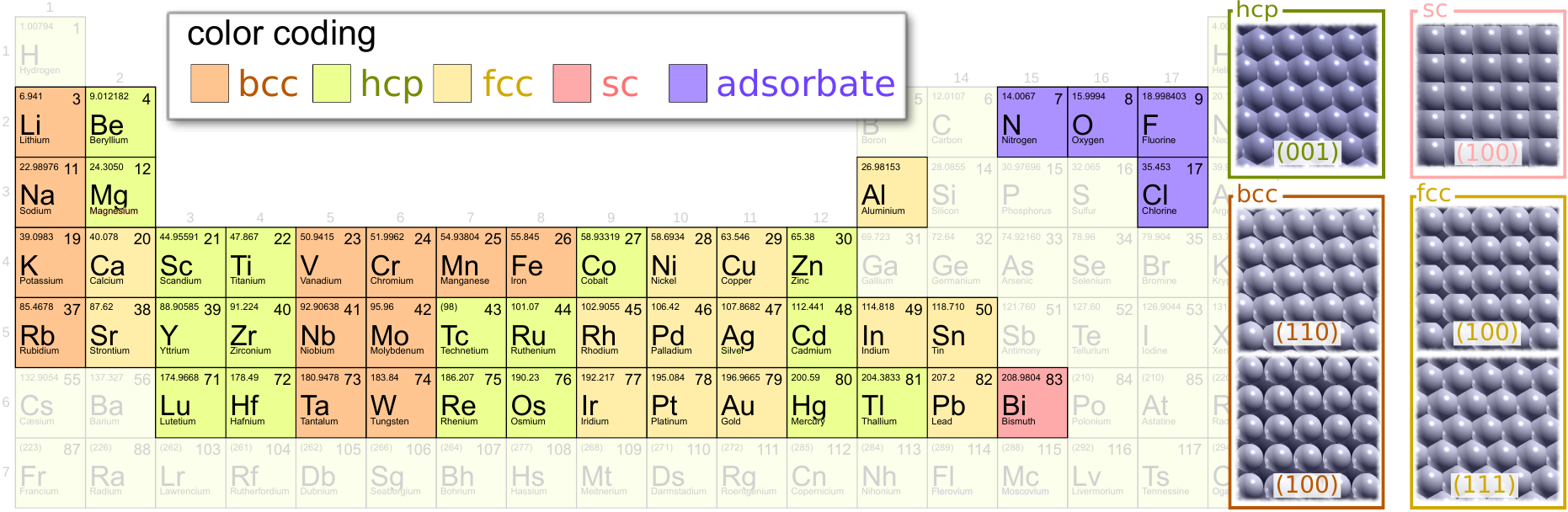}
   \caption[Investigated metallic surfaces and adsorbates]{ The
     investigated metals and adsorbates are highlighted in the
     periodic table. The lattice type of each investigated metal is
     indicated by the color coding; note that metals with exotic
     lattices were modeled as either bcc, fcc, hcp, or
       simple-cubic (sc) as explained in the text. The side panel
     displays topviews of the considered surfaces for each lattice
     type.}
   \label{fig:ChosenSystemsPeriodicTable}
\end{figure*}                                                                

Most of the investigated metals crystallize in one of the following
three lattice types: face-centered-cubic (fcc), hexagonal-close-packed
(hcp), and body-centered-cubic (bcc). The exceptions are In and Sn,
which crystallize in tetragonal lattices, as well as Hg and Bi, which
crystallize in rhombohedral lattices. For these metals the most stable
among fcc, hcp, bcc, or simple-cubic (sc) was chosen as the
representative model in order to simplify the
calculations. Additionally, $\alpha$-Mn has a unique bcc lattice with
58 atoms in the unit cell,\cite{Bradley_PRSLA115} however, for
simplicity we modeled it with a plain bcc lattice. The selected
Bravais lattice type for each investigated metal is indicated along
with the considered surfaces in
\Fig~\ref{fig:ChosenSystemsPeriodicTable}, i.e., (001) for hcp, (110)
and (100) for bcc, (100) and (111) for fcc, and (100) for sc
metals. In total, we considered 70 different surfaces. The adatoms
predominantly adsorb to hollow sites, although for some cases they
prefer top or bridge sites: these exceptions are listed in
\Tab~\ref{tab:exceptions} in the Supporting Information.

The adatom binding energy (\Eb), as defined by \Eq~\eqref{eq:Eb} in
the Supporting Information, was calculated for \obo\ and \tbt\ adatom
overlayers, designated as \Eobo and \Etbt, respectively. The
difference between the two binding energies ($\Del$):
\begin{equation}
  \label{eq:Del}
  \Del = \Eobo - \Etbt,
\end{equation}
was used as the criterion to determine whether lateral interactions
are attractive. As to differentiate between attractive (or
repulsive) and negligible lateral interactions, we arbitrarily adopt
a threshold of 0.1~eV and define interactions to be attractive if
$\Del < -0.1$~eV, negligible if
$-0.1~\mathrm{eV} \leq \Del \leq 0.1~\mathrm{eV}$, and repulsive
when $\Del > 0.1~\mathrm{eV}$.

\section*{RESULTS AND DISCUSSION}

\begin{figure*}[htb]
   \centering
   \includegraphics[width=0.94\textwidth]{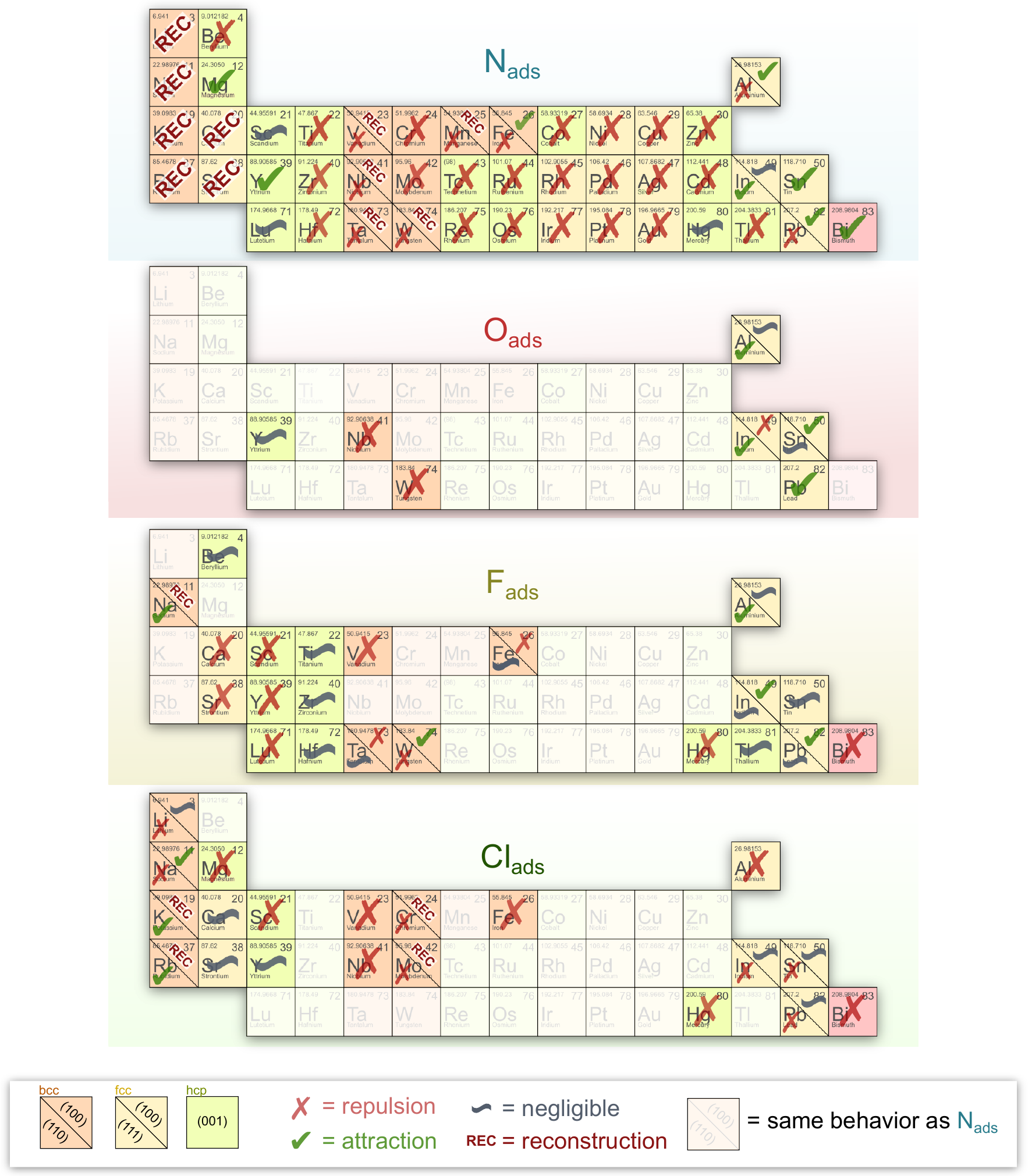}
   \caption[Summary of lateral interactions between adatoms on
   metallic surfaces]{A summary of lateral interactions between
     adatoms on investigated metal surfaces. Four scenarios were
     found: (i) repulsive interactions, (ii, iii) attractive or
     negligible interactions, and (iv) reconstruction. For the N
     adatom all results are explicitly summarized, whereas for other
     adatoms only differences with respect to the N adatom are
     shown. For bcc and fcc metals, the results for the two considered
     surfaces are shown as indicated by the legend.}
   \label{fig:AttractiveInteractionsResults}
\end{figure*}

\begin{figure}[ht]
   \centering
   \includegraphics[width=1.00\columnwidth]{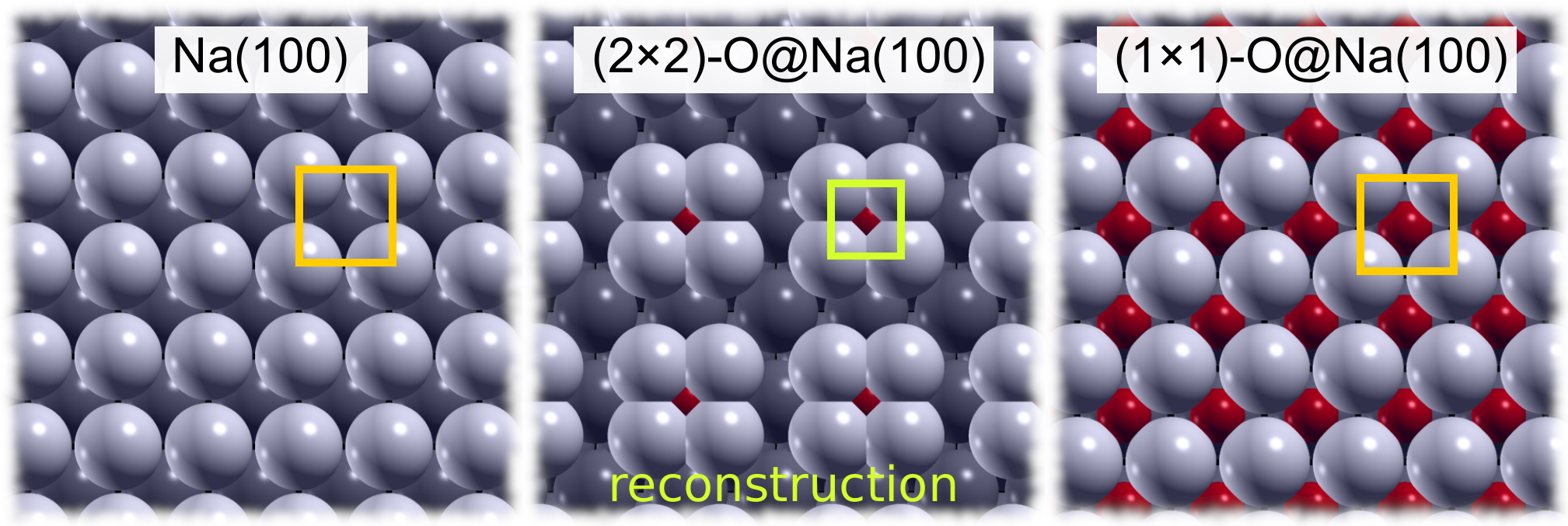}
   \caption[Example of surface reconstruction]{An example of surface
     reconstruction for O on Na(100). For the \tbt\ overlayer the Na
     atoms closest to the adatom move toward it so that Na$_4$O
     islands form. Such reconstruction is not possible for the \obo\
     overlayer due to symmetry. Such a reconstruction occurs for all
     metals labeled as ``REC'' in
     \Fig~\ref{fig:AttractiveInteractionsResults}, though the extent
     of reconstruction can vary considerably.}
   \label{fig:ReconstructionExample}
\end{figure}

The main result of this work is shown
\Fig~\ref{fig:AttractiveInteractionsResults}, which schematically
presents the type of lateral interactions for the N, O, F, and Cl
adatoms on 70 different surfaces of 44 elemental metals.
We find that lateral interactions can be classified into four
different groups: (i) the expected repulsive interactions; (ii,iii)
the case where the simple ionic model applies and the lateral
interactions are either attractive or negligible; and (iv) the case
where conditions of the simple ionic model are met, however, surface
reconstruction makes the low coverage \tbt\ overlayer more stable
than the high-coverage one.
Note that some cases belong to more than one scenario, nevertheless,
each specific case is described only by a single category. To
this end the following order of precedence is adopted: (1)
attractive interactions, (2) reconstruction, and (3) negligible or
repulsive interactions.
Reconstruction is characterized by metal atoms (ions) nearest to
the adatom being substantially displaced toward the adatom thus
forming island-like structures on the surface. A typical example is
shown in \Fig~\ref{fig:ReconstructionExample}.
  
In order to provide a quantitative measure of the extent of surface
reconstruction, we defined the reconstruction quotient (\RQ) as:
\begin{equation}
  \label{eq:RQ}
  \RQ = A_{\mathrm{R}}/A_{\obo}, 
\end{equation}
where $A_{\mathrm{R}}$ is the area of the reconstructed ``cell'' and
$A_{\obo}$ is the area of the \obo\ unit-cell (for a schematic
definition of these quantities see \Fig~\ref{fig:ReconDegree} in the
Supporting Information).
Because metal ions nearest to the adatom always respond to
its presence (by either moving toward or away from it), we define the
surface to be reconstructed only when the \RQ\ is significantly below
1; we arbitrarily set $\RQ \leq 0.9$ as the criterion for
reconstruction.

In addition to the aforementioned
\Fig~\ref{fig:AttractiveInteractionsResults}, which schematically
summarizes the results about lateral interactions, \Eb\ values for
each specific case are tabulated in
\Tabs~\ref{tab:Eb-bcc-100}--\ref{tab:Eb-sc-100} and plotted along with
\RQ\ values in
\Figs~\ref{fig:Eb-and-ReconDegree-bcc-100}--\ref{fig:Eb-and-ReconDegree-hcp-001}
in the Supporting Information.
In accordance with previous
studies,\cite{Miller_JCP134,Loffreda_JCP108,Gava_PRB78,Ma_SS619,Peljhan_JPCC113,Inderwildi_JCP122,Gossenberger_SS631}
our results reveal that repulsion is the dominant case for
electronegative adatoms on d-block metal surfaces with few exceptions,
such as Fe(100), on which N and O adatoms display attractive
interactions, and Hg(001), which displays either attractive,
negligible, or repulsive interactions; negligible lateral interactions
were also identified for some adatoms on group 3 and 4 d-block hcp
metals.
Additionally, reconstruction occurs for (100) surfaces of several bcc
d-block metals.

Attractive or negligible lateral interactions are the dominant
scenario on the surfaces of p-block metals. In particular N, O, and F
display such behavior on a large majority of investigated p-block
metal surfaces. Exceptions are repulsive interactions for N on
Al(111), Tl(001), and Rb(111); O on In(100) and Tl(001); and F on
Bi(100). In contrast, Cl mainly displays repulsive lateral
interactions on p-block metal surfaces, with the exception of In(100),
Sn(100), and Pb(100) where lateral interactions are negligible.

The third group are the s-block metals where the dominant scenario is
reconstruction, in particular for N, O, and to lesser extent for F
adatoms. Notable exceptions are Mg and Be, where lateral interactions
are attractive and repulsive, respectively. In contrast, for Cl
reconstruction occurs only on K(100) and Rb(100), whereas on other
surfaces of s-block metals Cl generally displays either attractive or
negligible lateral interactions, except on Li(110), Be, Na(110), and
Mg where the interactions are repulsive.

\begin{figure}[tb]
  \centering
  \includegraphics[width=1.0\columnwidth]{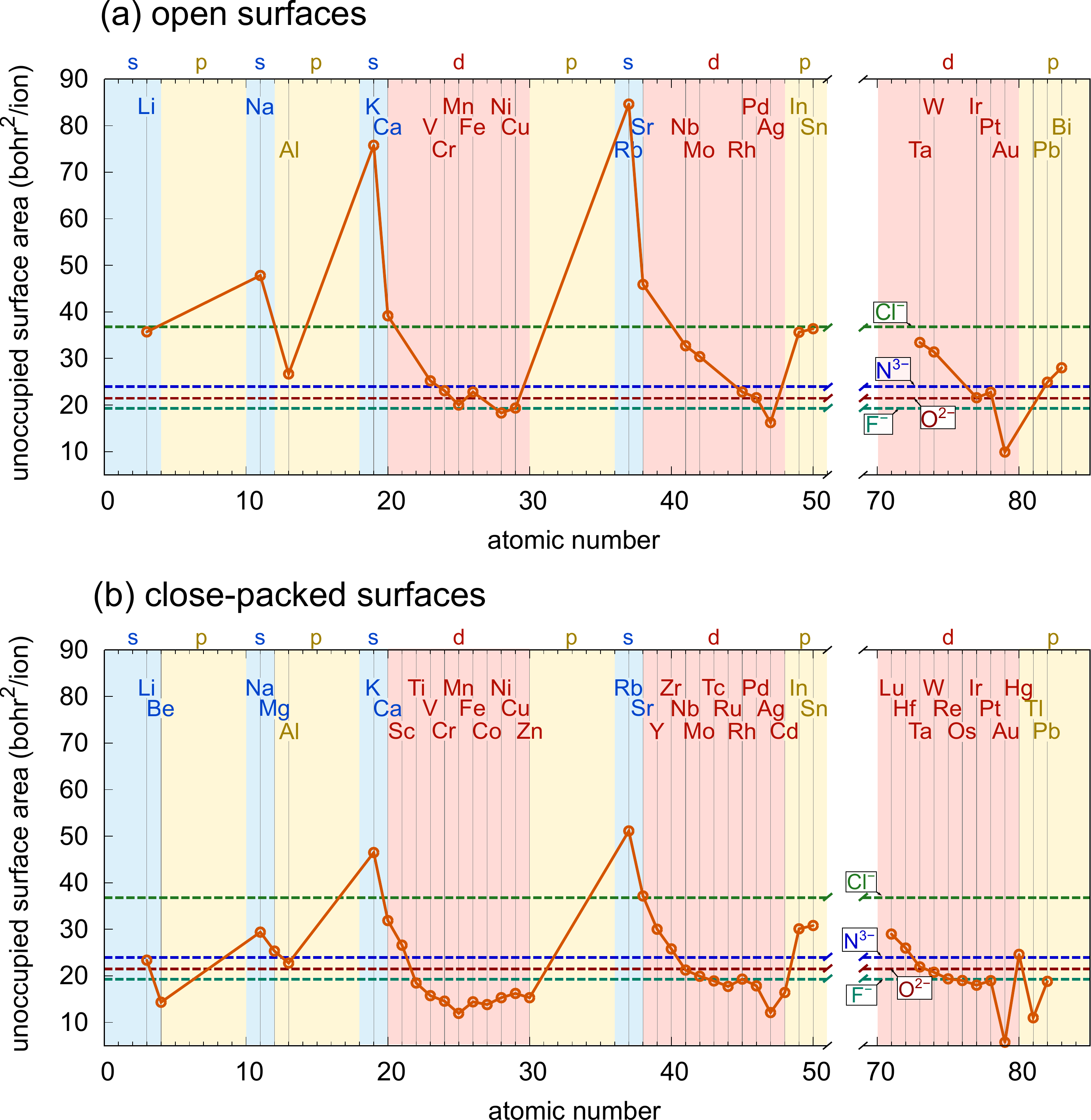}
  \caption[Results of unoccupied surface area analysis]{Unoccupied
    surface area (\Aex) for (a) open surfaces [fcc(100), bcc(100), and
    sc(100)] and (b) close-packed surfaces [fcc(111), hcp(001), and
    bcc(110)]. The horizontal dashed lines indicate the \Aa\ of
    adsorbed adatoms, calculated as described in the text.}
  \label{fig:Aex}
\end{figure}

Our results indicate that in some cases, such as N, O, and F on alkali
metals, where the two conditions of the simple ionic model for the
attractive lateral interactions are met (ionic adatom--surface bonding
and low height of adatoms), reconstruction occurs instead. This
implies that the simple ionic model cannot describe all the situations
and needs to be extended, at least conceptually, as to account for the
possibility of reconstruction. To this end, we define two quantities
termed {\it unoccupied surface area} (\Aex) and {\it area occupied by
  the anion} (\Aa), defined as:
\begin{equation}
  \label{eq:ExcessArea}
  \Aex = A_{\obo} - \pi R^2_{\mathrm{c}} \quad \text{and}
  \quad
  \Aa = \pi R^2_{\mathrm{a}},
\end{equation}
where $A_{\obo}$ is the area of the \obo\ surface cell, \Rc\ is the
ionic radius of the metal cation, calculated as the average of the
effective ionic radii for all coordination numbers of the metal in the
lowest cationic oxidation state,\cite{Shannon_ACB26} and \Ra\ is the
effective radius of the anion \cite{Shannon_ACB26} (for a graphical
representation of the {\it unoccupied surface area}, see
\Fig~\ref{fig:VacantArea} in the Supporting Information).
The comparison between \Aex\ and \Aa\ is presented in
\Fig~\ref{fig:Aex}.  This figure reveals that alkali metals, Ca, and
Sr display the largest \Aex\ and reconstructions typically occur on
their surfaces, in particular for N and O adatoms. Furthermore, it is
also evident from the figure that \Aa\ of Cl$^-$ is much larger than
\Aa\ of the other three adatoms and for this reason reconstructions
and attractive interactions are considerably less frequent for Cl
adatoms (cf.~\Fig~\ref{fig:AttractiveInteractionsResults}). The next
relevant observation is that repulsive interactions usually
appear when \Aex\ is small, i.e., when $\Aex\lesssim\Aa$. This is the
case of transition metal surfaces, where repulsive interactions
dominate.
Finally, if neither $\Aex\gg\Aa$ nor $\Aex\lesssim\Aa$ applies, then
the interactions are likely attractive or negligible. There are, of
course, exceptions, because such a simple rule simply cannot
encompass all cases.

\begin{figure}[tb]
  \centering
  \includegraphics[width=1.00\columnwidth]{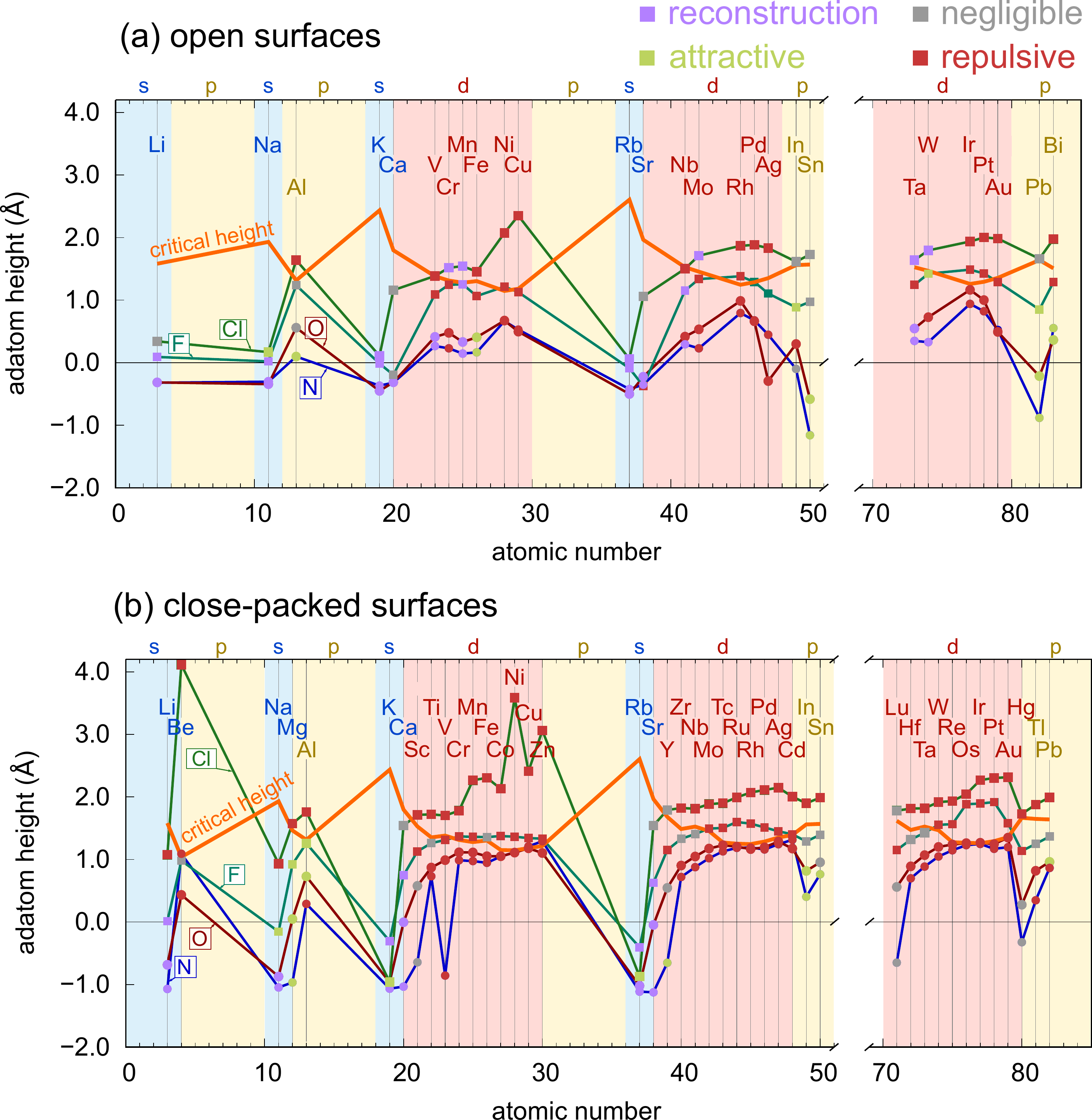}
  \caption[Critical-height analysis]{Comparison between critical
    heights as predicted by the simple ionic model and DFT calculated
    adatom heights for (a) open surfaces [fcc(100), bcc(100), and
    sc(100)] and (b) close-packed surfaces [fcc(111), hcp(001), and
    bcc(110)]. Each datapoint is color coded according to the type of
    interaction as indicated by the legend at the top right.}
  \label{fig:critical-height}
\end{figure}

As a further scrutiny of the utility of the simple ionic model,
let us compare the critical height above the surface---i.e., the
height above which the simple ionic model predicts that lateral
interactions between adatoms are repulsive---with the adatoms
heights predicted by DFT calculations
(\Fig~\ref{fig:critical-height}).
Notably, there is not a single case of attractive lateral interactions
with the adatoms located above the critical height. This observation
is very reassuring and provides strong support to the validity of our
explanation based on the simple ionic model, which differs from the
explanations provided by Jacobsen et al.\cite{Jacobsen_PRB52} for O on
Al(111) and by Cheng et al.\cite{Cheng_PRL113} for N, O, and F adatoms
on Mg(0001).  Jacobsen et al.\cite{Jacobsen_PRB52} emphasized the role
of Al p-states that open new possibilities for hybridization and
consequently lead to stronger bonding configurations at high coverage,
but this explanation is brought into question
by the aforementioned attractive interactions of N, O, and F adatoms
on Mg(0001),\cite{Cheng_PRL113} which is an s-metal, as well as by the
current findings---among the 24 identified cases of attractive
interactions, 11 of them appear on non p-metals (7 on s- and 4 on
d-metals, cf.~\Fig~\ref{fig:AttractiveInteractionsResults}).
The attractive interactions on Mg are accompanied by an adsorption
induced decrease of the work function, which is another anomaly
because an increase is typically expected for electronegative
adatoms.\cite{Michaelides_PRL90} Cheng et al.\cite{Cheng_PRL113}
attributed both anomalies to a highly polarizable electron spill-out
in front of Mg(0001), i.e., the vertical electron charge
redistribution (a depletion of charge above the adatom) causes the
decrease of the work function,\cite{Cheng_PRL113,Michaelides_PRL90}
whereas attractive interactions were explained by quantum mechanical
screening, i.e., a lateral transfer of the spill-out
electrons.\cite{Cheng_PRL113} In contrast, our explanation involves
neither the metal p-states nor the highly polarizable electron
spill-out, but instead explains the attraction by the simple ionic
model---i.e., an interplay of electrostatic and geometric
effects---requiring only unpolarizable point ions.
It is worth noting that on Mg(0001) the attractive lateral
interactions are indeed accompanied by an adsorption induced decrease
of the work function, however, the latter is not required for
attraction to emerge, as evidenced by \Fig~\ref{fig:dWf-vs-dEb} in the
Supporting Information, which shows the adsorption induced work
function change for all currently identified ``attraction cases''
(whereas \Fig~\ref{fig:dWf-all} shows work function changes for all
considered overlayers).
Among 24 such cases, the work function reduces for only 10 of them.
%

Turning back to \Fig~\ref{fig:critical-height}, its scrutiny for s-
and p-block metals reveals
that when the adatom height is below the critical height, the lateral
interactions are either attractive, negligible, or the surface
reconstructs. There are only a few exceptions, i.e., F on Bi(100), O
on Tl(001), and N on Tl(001) and Pb(111). The situation is
considerably different for transition metals, because for many cases
the adatoms are below the critical height, yet the lateral
interactions are repulsive. However, transition metals do not fulfill
the second requirement of the simple ionic model, that is, the
adatom--surface bonding is not sufficiently ionic, due to significant
participation of covalent
bonding.\cite{Peljhan_JPCC113,Baker_JACS130,Migani_JPCB110,Roman_PCCP16}
Note that transition metals are rather electronegative with
work-function values typically above 4~eV\cite{Michaelson_JAP48} (see
also \Fig~\ref{fig:work-function} in the Supporting Information);
exceptions are group-3 metals, which display lower work-functions, but
thereon the lateral interactions are usually not predicted by DFT to
be repulsive.

Finally, let us focus in more detail on
cases denoted as ``reconstruction'', where the lower coverage \tbt\
adatom overlayer is more stable than the high coverage \obo\
overlayer. Our analysis indeed reveals that the superior stability of
the \tbt\ overlayer is by and large due to reconstruction, where the
metal ions nearest to the adatom move laterally toward it, forming
island-like structures on the surface
(cf.~\Fig~\ref{fig:ReconstructionExample}).
For example, O on Na(100) displays a \Del\ of 1.8~eV. However, if the
larger Cl$^-$ ion is adsorbed on Na(100), reconstruction is no
longer viable and attractive interactions are found with a \Del\ of
  $-0.2$~eV.

The extent to which reconstruction stabilizes the \tbt\ overlayer of O
on Na(100) was estimated by performing a constrained relaxation, where
the lateral coordinates of Na atoms in the topmost layer were
constrained to their bulk positions.
The resulting \Del\ reduces from 1.8~eV for the reconstructed
structure to 0.2~eV for the constrained structure, which implies that
reconstruction stabilizes the \tbt\ overlayer by 1.6~eV, which is
considerable.  Notice, however, that even without reconstruction, the
\tbt\ overlayer remains slightly more stable.  The reason for the
superiority of the \tbt\ overlayer can be attributed to the large
lattice constant of Na that diminishes the magnitude of electrostatic
stabilization (the effect is illustrated in
\Fig~\ref{fig:ionic_model_predicitions} of the Supporting
Information).  Thus the lack of attractive interactions, even when the
top layer is constrained, is likely a consequence of diminished
stabilization in combination with other effects, not taken into
account by the simple ionic model.

\section*{CONCLUSION}  

To summarize, by performing DFT calculations of the adsorption of four
different electronegative adatoms on 70 surfaces of 44 elemental
metals, we showed that even something as conceptually simple as
adsorption of electronegative adatoms on metal surfaces, can lead to
unanticipated behavior.
Understanding such interactions is important for heterogeneous
catalysis and electrochemistry as they may provide a new insight into
initial stages of corrosion and passivation.
We identified four possible scenarios for the lateral interactions
between electronegative adatoms, some of them being unexpected, and
explained the reasons why they emerge. Lateral interactions can be:
(i) repulsive (this is the expected scenario, but it prevails only on
d-block metals), (ii, iii) attractive or negligible (this scenario is
predominantly found for p-block metals and Mg; their origin is well
explained by our simple ionic model, i.e., attraction is a consequence
of predominantly ionic bonding and a low height of adatoms above the
surface), or (iv) surface reconstruction of the lower coverage \tbt\
overlayer provides additional stabilization, making it more stable
than the high-coverage \obo\ overlayer. This case typically occurs on
s-block metals.

\section*{ACKNOWLEDGEMENT}

This work has been supported by the Slovenian Research Agency
(Grants No. P2-0393).

\bibliography{biblio,footnotes}

\clearpage
\onecolumn
\vspace*{3em}
\centering
\section*{\LARGE For Table of Contents Only}
\vspace*{6em}
\includegraphics[width=8.25cm]{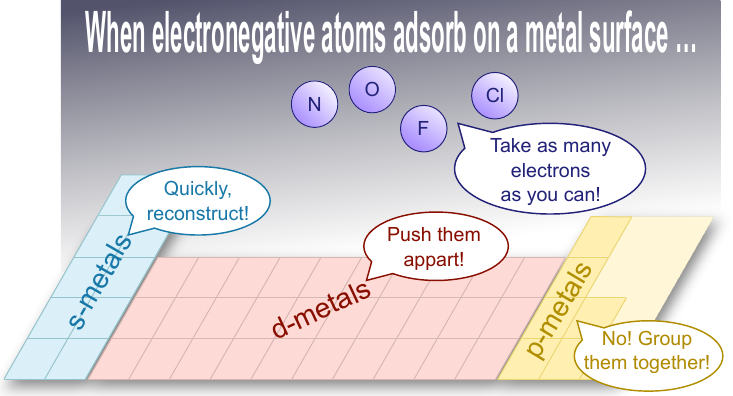}

\end{document}


\title{\textbf{Surprising lateral interactions between negatively charged adatoms on metal surfaces\\[1em]{\LARGE\bf SUPPORTING INFORMATION}}}

\author{Matic Pober\v znik and Anton Kokalj*\\[1em]
  \normalsize
  Department of Physical and Organic Chemistry,
  Jo\v zef Stefan Institute, Jamova 39, SI-1000 Ljubljana, Slovenia
}

\date{\small\today}

\maketitle

\thanks{\noindent
  \rule{0.2\textwidth}{1pt}\\
  \footnotesize
  ORCID IDs: 0000-0002-4866-4346 (MP), 0000-0001-7237-0041 (AK)\\
  * Corresponding Author: Anton Kokalj, Tel: +386-1-477-35-23; 
  Fax: +386 1 251 93 85, E-mail: tone.kokalj@ijs.si,
  URL: \texttt{http://www.ijs.si/ijsw/K3-en/Kokalj}\\
}

\section{Description of the simple ionic model}

\begin{figure}[t]
  \centering
  \includegraphics[width=1.0\columnwidth]{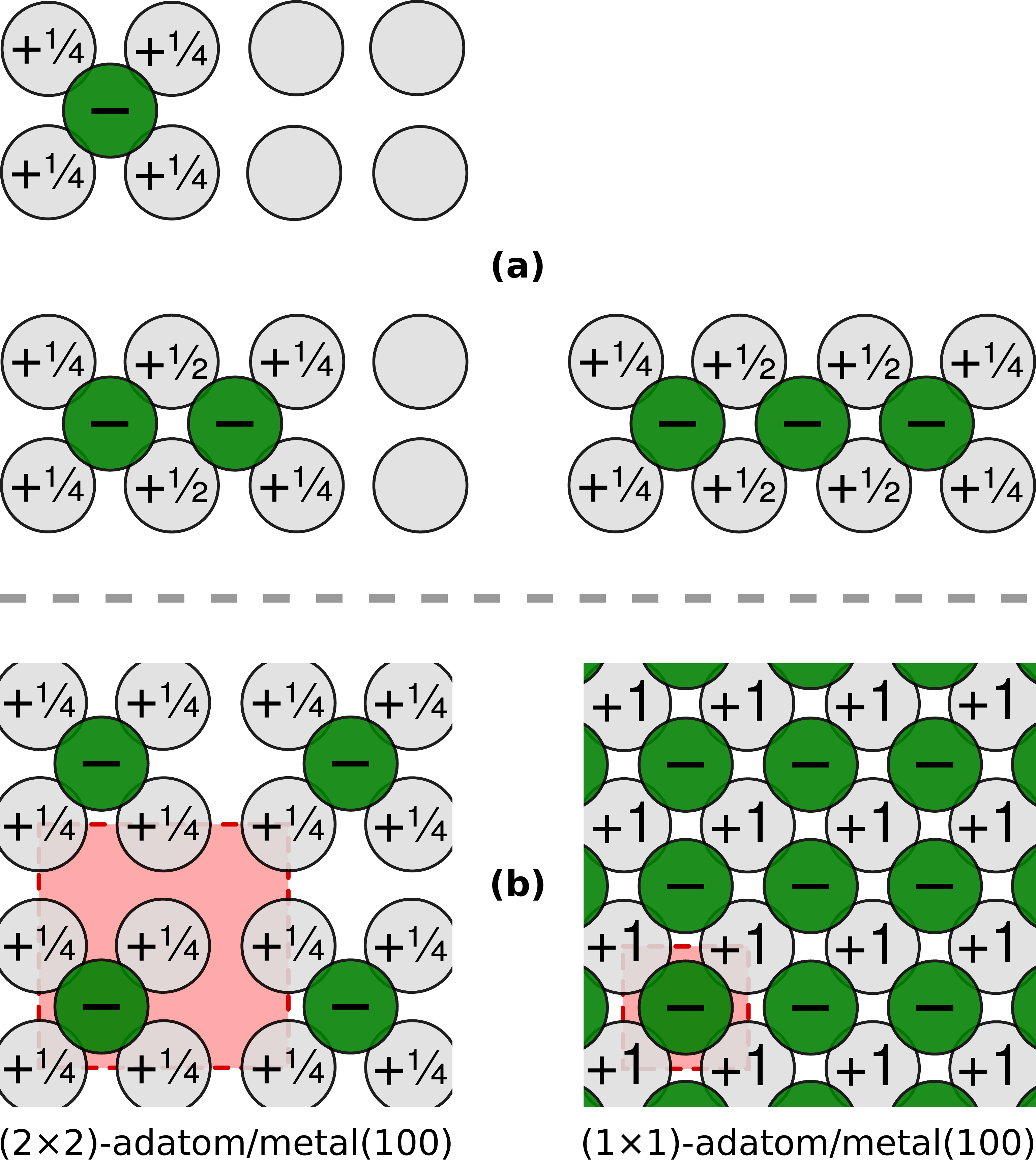}
  \caption{Schematic of the electron charge distribution among adatoms
    and nearest neighbor metal atoms in the simple ionic model. (a)
    The charge on adatoms is set to \qad\ (labeled as ``$-$''),
    whereas the counter-charge is distributed proportionally to the
    nearest neighbor metal atoms, i.e., for an adatom with $n$ metal
    neighbors each of them donates $\qad/n$ electrons to the
    adatom. If a metal atom has no adatom neighbors then it remains
    charge-neutral, but if it has $m$ adatom neighbors then it donates
    $\qad/n$ electrons to each, thus in total $m\qad/n$ to all of
    them. (b) Charge distribution for \tbt\ and \obo\ adatom
    overlayers on a square lattice of metal atoms, which is compatible
    with bcc(100), fcc(100), and sc(100). Unit-cells are indicated in
    red.}
   \label{fig:ionic-model-charge}
\end{figure}

The ``simple ionic model'' was derived in our previous
publication\cite{Poberznik_JPCC120} and here we briefly explain its
essence.
The model is based on the electronic structure analysis of the
O/Al(111) system,\cite{Poberznik_JPCC120} which reveals that O--Al
bonding is ionic and furthermore that the excess electron charge on
adatoms mainly comes from the nearest neighbor metal
atoms.\cite{Poberznik_JPCC120}
%
%
The last observation is exploited in the simple ionic model, where the
adatom gets the electron charge exclusively from the nearest neighbor
metal atoms, such that each neighboring metal atom contributes
proportionally, as schematically shown in
\Fig~\ref{fig:ionic-model-charge}. The simple ionic model therefore
consists of an ionic bilayer of adatom-anions/metal-cations and can be
described by $N$ ions in the unit-cell at positions
$\{\bm{\tau}_i\}_{i=1}^N$ and charges $\{q_i\}_{i=1}^N$. The
interaction energy is then obtained by summing the pairwise Coluomb
interactions among the ions in the infinite adatom/metal
bilayer, i.e.:
\begin{align}
  \label{eq:Eint_general}
  E_{\rm int} =&
                 \frac{1}{2}\sum_{\vR=\bm{0}}^{\infty}\sum_{i,j}^{N}
                 \frac{q_iq_j}{|\vR + \bm{\tau}_j - \bm{\tau}_i|}
                 (1-\delta_{i,j}\delta_{R,0}),
\end{align}
where \{\vR\} are the lattice vectors of a two-dimensional
lattice. The role of the $(1-\delta_{i,j}\delta_{R,0})$ term is to
omit the interaction of an ion with itself ($i=j$ and $R=0$, where
$R=|\vR|$). The infinite lattice sum in two-dimensions,
$\sum_{\vR\ne0}^\infty\left(\cdots\right)$, can be evaluated by
explicitly calculating it within the cutoff radius \Rcut, whereas
beyond \Rcut\ it is approximated by an integral, in particular:
\begin{align}
  \label{eq:Eint_int}
  \nonumber
    E_{\rm int}
    \simeq& \
    \frac{1}{2}\left(\sum_{\vR=\bm{0}}^{|\vR|<\Rcut}\sum_{i,j}^{N}
      \frac{q_iq_j}{|\vR + \bm{\tau}_j - \bm{\tau}_i|}
      (1-\delta_{i,j}\delta_{R,0})\right.\\
  \nonumber
          &\left. \quad\quad\quad +\ \frac{2\pi}{A}\int_{\Rcut}^\infty
            \sum_{i,j}^{N}\frac{Rq_iq_j}{\sqrt{R^2 + (z_j -
            z_i)^2}}{\rm d}R\right)\\
  \nonumber
  =&\ \frac{1}{2}\left(\sum_{\vR=\bm{0}}^{|\vR|<\Rcut}\sum_{i,j}^{N}
     \frac{q_iq_j}{|\vR + \bm{\tau}_j - \bm{\tau}_i|}
     (1-\delta_{i,j}\delta_{R,0})\right.\\
          &\left. \quad\quad\quad -\ 
            \frac{2\pi}{A}\sum_{i,j}^{N}
            q_iq_j\sqrt{R_{\rm cut}^2+ (z_j - z_i)^2}\right),
\end{align}
where $A$ is the area of the unit-cell and $z_i$ is the
$\hat{z}$-coordinate of the atomic position $\bm{\tau}_i$, i.e.,
$\bm{\tau}_i = (x_i, y_i, z_i)$.
Note that due to charge neutrality,
the larger is the \Rcut, the smaller is the last sum
($-\frac{2\pi}{A}\sum_{i,j}^{N}\cdots$), due to cancellation between
its terms. In particular, the last sum scales as
$(2\pi/A)\mu_z^2R_{\rm rcut}^{-1}$,
where $\mu_z$ is the $\hat{z}$-component of the dipole of the
ions in the unit-cell, $\mu_z = \sum_{i=1}^Nz_iq_i$. Hence:
\begin{equation}
  \label{eq:charge-dipole}
  -\sum_{i,j}^{N}
  q_iq_j\sqrt{R_{\rm cut}^2+ (z_j - z_i)^2}
  \simeq \mu_z^2R_{\rm rcut}^{-1}.
\end{equation}

\begin{figure}[t]
   \centering
   \includegraphics[width=1.0\columnwidth]{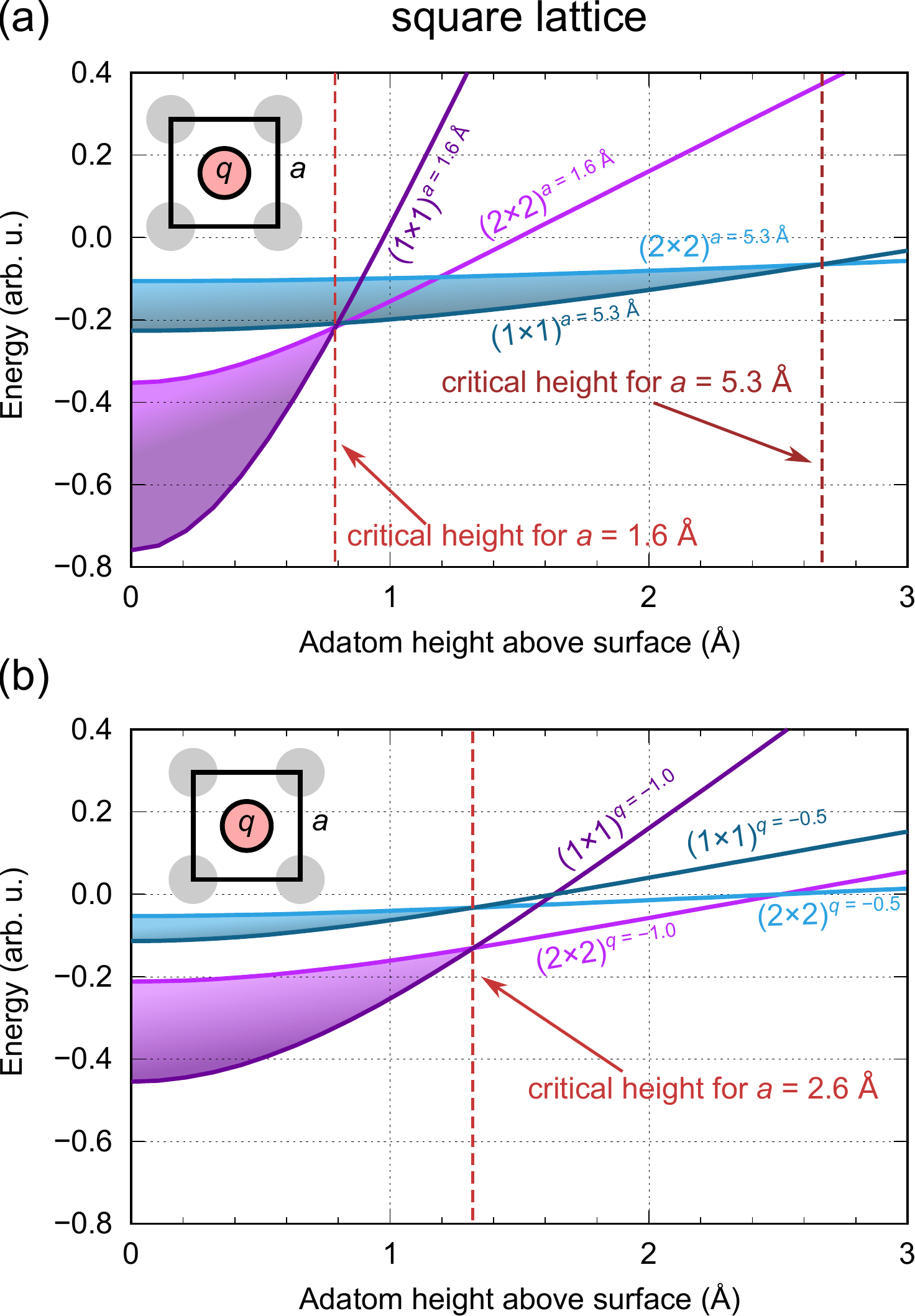}
   \caption{The prediction of the simple ionic model for the \obo\ and
     \tbt\ adatom overlayers above a square lattice of metal
     atoms, which is compatible with the bcc(100), fcc(100), and
     sc(100) surfaces.  %
     Notice that below the critical height the high-coverage \obo\
     overlayer is more stable than the lower-coverage \tbt\
     overlayer. (a) The dependence of the interaction energy and the
     critical height on the lattice parameter for $a=1.6$~\AA\ and
     $a=5.3$~\AA.
     Notice that the \obo\ over \tbt\ preference below the critical
     height decreases by increasing the lattice parameter, while
     concomitantly the critical height becomes larger. (b) The
     dependence of the interaction energy on the charge of the adatom
     for $q=-0.5$ and $q=-1.0$. Because the energy depends
     quadratically on the charge, the \obo\ over \tbt\ preference
     below the critical height decreases with a decrease in charge
     magnitude.}
   \label{fig:ionic_model_predicitions}
\end{figure}                                                                

\Fig~\ref{fig:ionic_model_predicitions} depicts the results of the
simple ionic model for the \obo\ and \tbt\ adatom overlayers over a
square lattice of metal atoms, with the interaction energy shown as a
function of adatom height above the surface. The figure illustrates
the dependence of the interaction energy on the lattice parameter
(\Fig~\ref{fig:ionic_model_predicitions}a) and on the adatom charge
(\Fig~\ref{fig:ionic_model_predicitions}b). The most important result
of the ionic model is that there exist a critical adatom height below
which the high-coverage \obo\ overlayer is more stable than the low
coverage overlayers (currently only the \tbt\ overlayer is considered
for low coverage, but in our previous work\cite{Poberznik_JPCC120} we
considered even lower coverages). This effect is referred to as
``stabilization'' in the following. The critical height obviously
depends on the lattice parameter due to geometric reasons, i.e., the
larger is the lattice parameter, the larger is the critical height.
Concomitantly with the increase of the lattice parameter the extent of
stabilization obviously decreases
(\Fig~\ref{fig:ionic_model_predicitions}a).  As for the dependence on
the adatom charge, the stabilization obviously increases with the
magnitude of the adatom charge
(\Fig~\ref{fig:ionic_model_predicitions}b), because the energy depends
quadratically on the charge.
Some of these dependencies can be easily understood from
\Eq~\eqref{eq:Eint_general}.

\section{Definitions}

\subsection{Binding energy}

DFT calculated binding energies were calculated as:
\begin{equation}
  \label{eq:Eb}
\Eb = \Esub{\mathit{X}/slab} - \Esub{X} - \Esub{slab},
\end{equation}
where $X$ stands for the adatom (either N, O, F, or Cl) and
$\Esub{\mathit{X}/slab}$, $\Esub{X}$, and $\Esub{slab}$ are the total
energies of the adatom-slab system, standalone adatom, and bare slab,
respectively. The binding energies for bcc, fcc, and hcp adatom/metal
systems are plotted in \Figs~\ref{fig:Eb-and-ReconDegree-bcc-100} to~\ref{fig:Eb-and-ReconDegree-hcp-001} and tabulated in
\Tabs~\ref{tab:Eb-bcc-100} to \ref{tab:Eb-hcp-001-hcpsite}.  The \Eb\
values of sc (simple-cubic) systems are given in
\Tab~\ref{tab:Eb-sc-100}, but they are not plotted, because only
Bi is ``considered'' to crystallize in this lattice type.
Note that Bi and some other investigated metals crystallize in more
``exotic'' lattices, however, in order to simplify the calculations,
they were modeled by one among the bcc, fcc, hcp, or sc lattice types,
as described in the main text.
%

\begin{figure}[h]
   \centering
   \includegraphics[width=1.0\columnwidth]{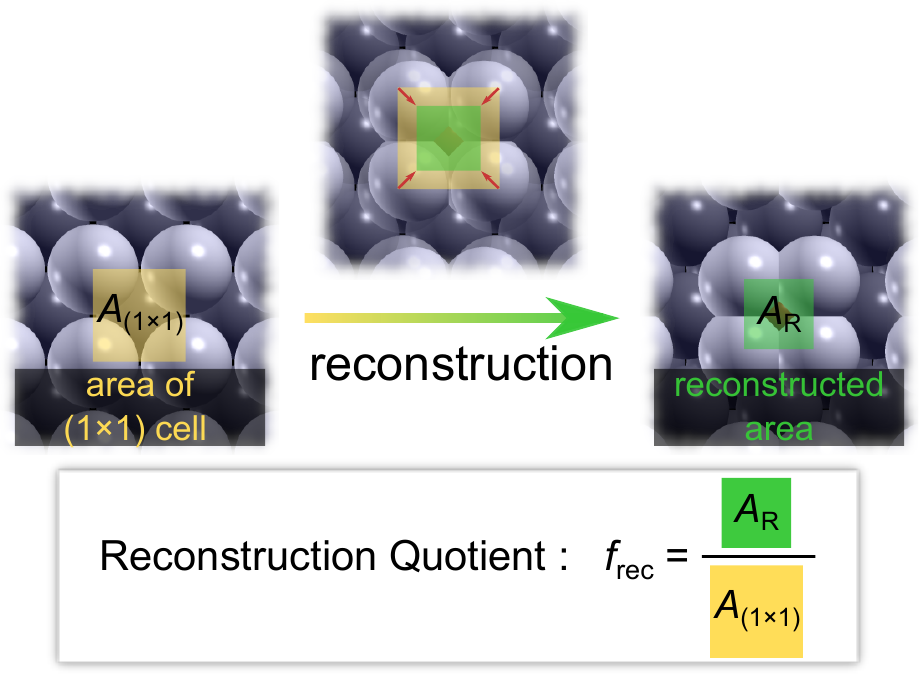}
   \caption{A schematic definition of the reconstruction quotient,
     $\RQ = A_{\mathrm{R}}/A_{\obo}$. $A_{\obo}$ is the area of the
     \obo\ unit-cell (left panel), whereas $A_{\mathrm{R}}$ is the
     area enclosed by the four metal atoms nearest to the adatom
     forming a \tbt\ overlayer (right panel). The reconstruction
     quotient was used to estimate the degree of reconstruction, in
     particular, we used $\RQ \leq 0.9$ as the criterion for
     reconstruction.}
   \label{fig:ReconDegree}
\end{figure}                                                                

\subsection{Reconstruction quotient}

In addition to attractive and repulsive lateral interactions we also
identified another possibility, when the two conditions of the simple
ionic model for the attractive lateral interactions are met (ionic
adatom--surface bonding and low height of adatoms). In particular,
surface reconstruction stabilizes the \tbt\ overlayer and makes it
more stable than the \obo\ overlayer. In order to quantify the extent
of reconstruction for each adatom/metal pair, we defined the
reconstruction quotient (\RQ) by \Eq~\eqref{eq:RQ} in the main
text. The way the reconstruction quotient is calculated is
schematically illustrated in \Fig~\ref{fig:ReconDegree}.  We defined
the surface to be reconstructed only when \RQ\ is significantly below
1; we arbitrarily set $\RQ \leq 0.9$ as the criterion for
reconstruction.

We should comment on how the reconstruction quotient was calculated
for close-packed bcc(110), fcc(111), and hcp(001) surfaces.  In
particular, on bcc(110) the adatoms were found in two distinct sites,
i.e., three-fold and four-fold hollow sites (see
\Fig~\ref{fig:bcc-110-sites}), whereas on fcc(111) and hcp(001) both
fcc- and hcp-hollow sites are three-fold coordinated.
For all the three-fold hollow sites, \RQ\ was estimated by considering
the area spanned by the three metal cations nearest to the adatom,
whereas for the bcc(110) four-fold hollow site the area spanned by the
four nearest metal cations was taken into account.

\begin{figure*}[hb]
   \centering
   \includegraphics[width=.95\textwidth]{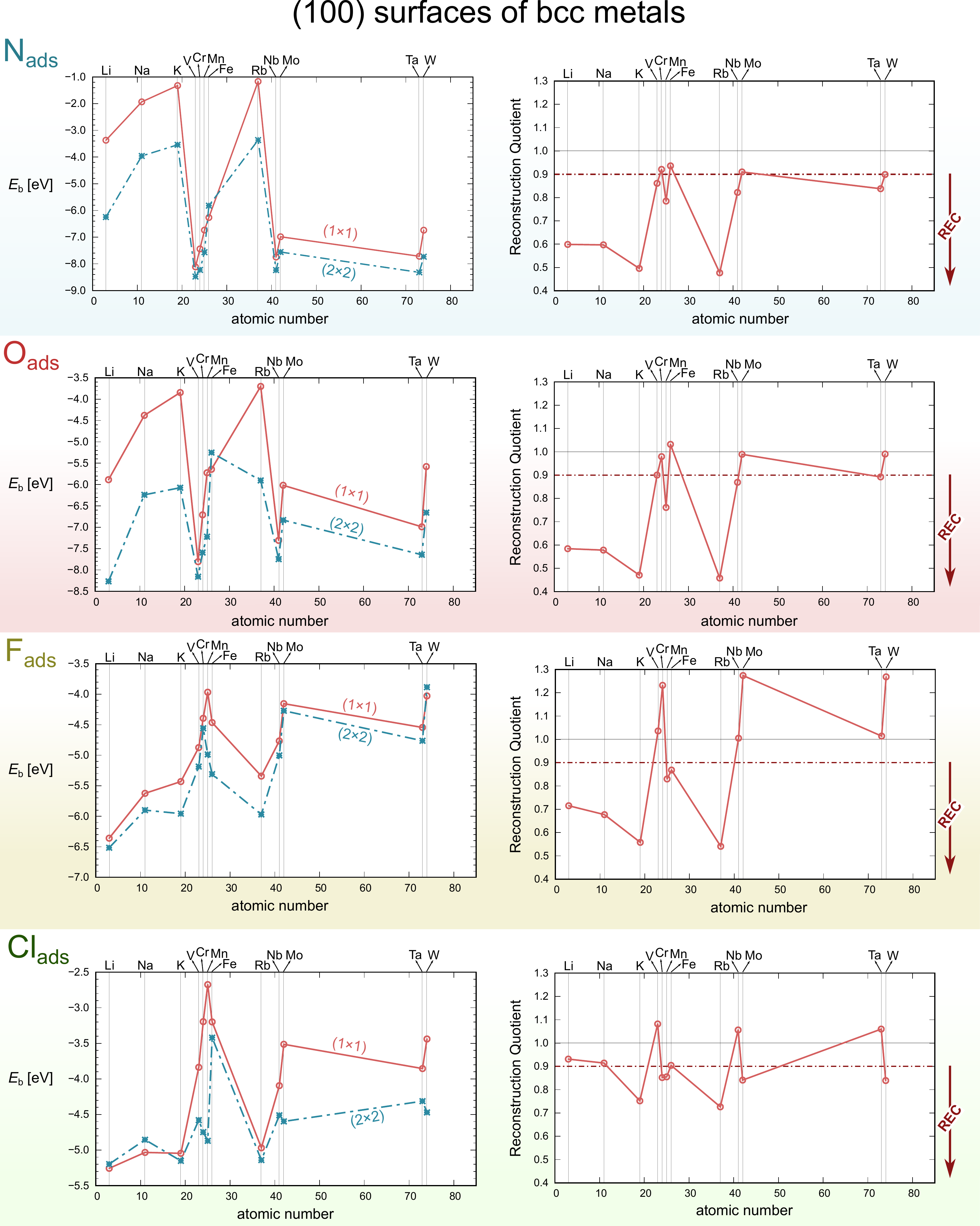}
   \caption{Left: binding energies (\Eb) of adatoms on
     bcc(100) surfaces for \obo\ and \tbt\ overlayers. Right: the
     degree of reconstruction, expressed in terms of the
     reconstruction quotient (\RQ).
     The red dash-dotted line at $\RQ=0.9$ indicates the adopted
     threshold for the reconstruction. Lines are drawn to guide the
     eye.}
   \label{fig:Eb-and-ReconDegree-bcc-100}
\end{figure*}
\begin{figure*}[htp]
   \centering
   \includegraphics[width=0.97\textwidth]{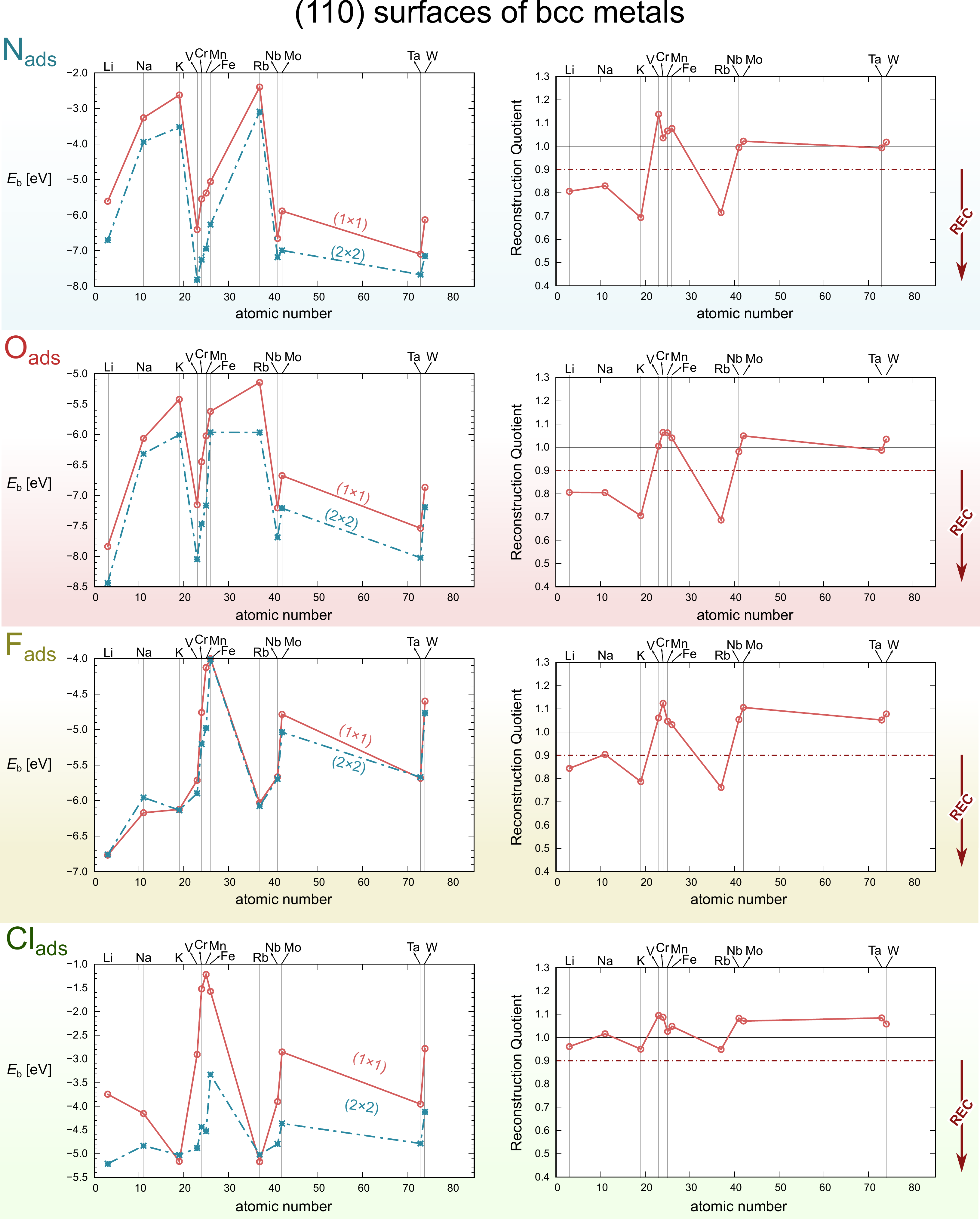}
   \caption{As in \Fig~\ref{fig:Eb-and-ReconDegree-bcc-100}, but
     for bcc(110) surfaces.}
   \label{fig:Eb-and-ReconDegree-bcc-110}
\end{figure*}
\begin{figure*}[htp]
   \centering
   \includegraphics[width=0.97\textwidth]{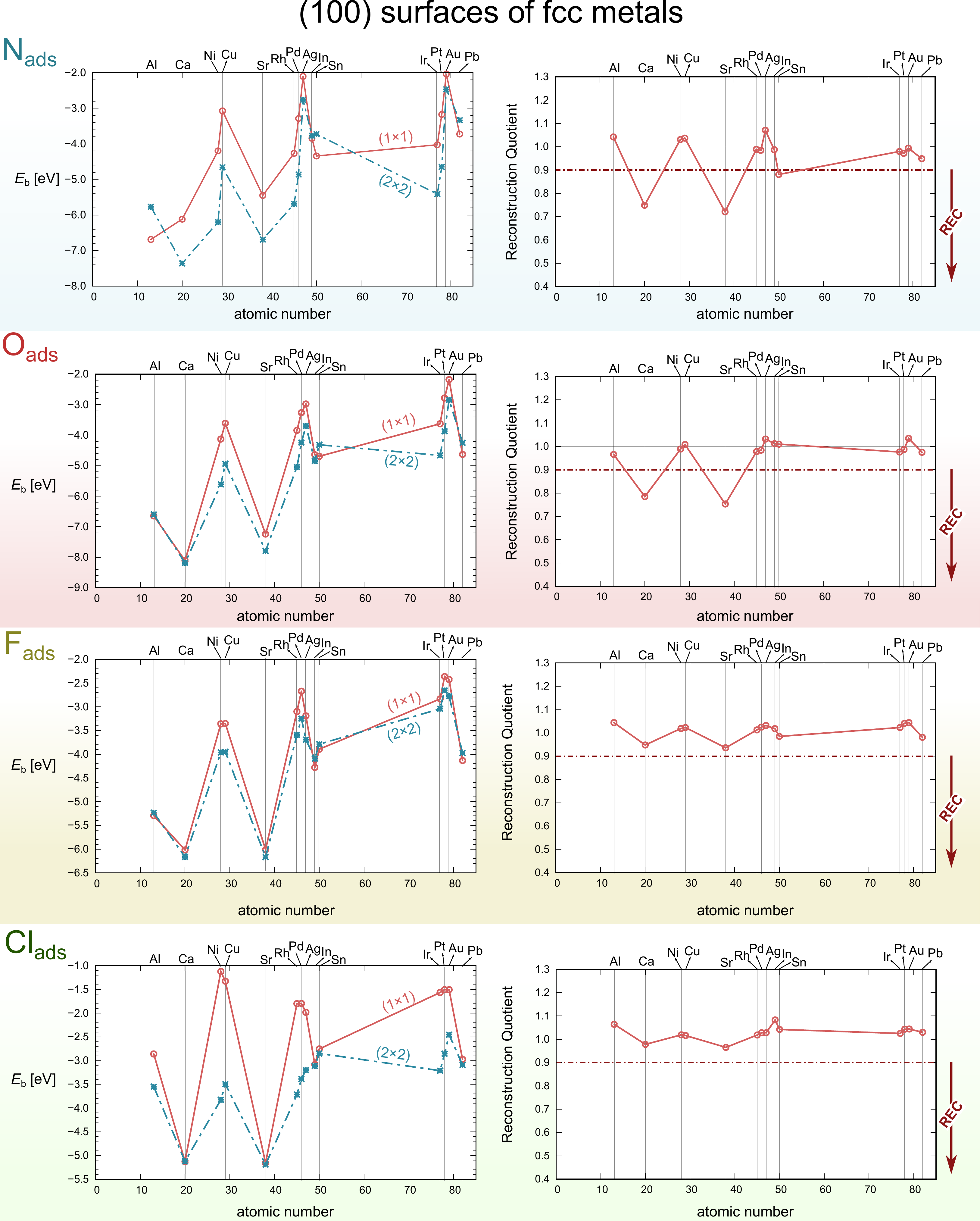}
   \caption{As in \Fig~\ref{fig:Eb-and-ReconDegree-bcc-100}, but
       for fcc(100) surfaces.}
   \label{fig:Eb-and-ReconDegree-fcc-100}
\end{figure*}
\begin{figure*}[htp]
   \centering
   \includegraphics[width=0.97\textwidth]{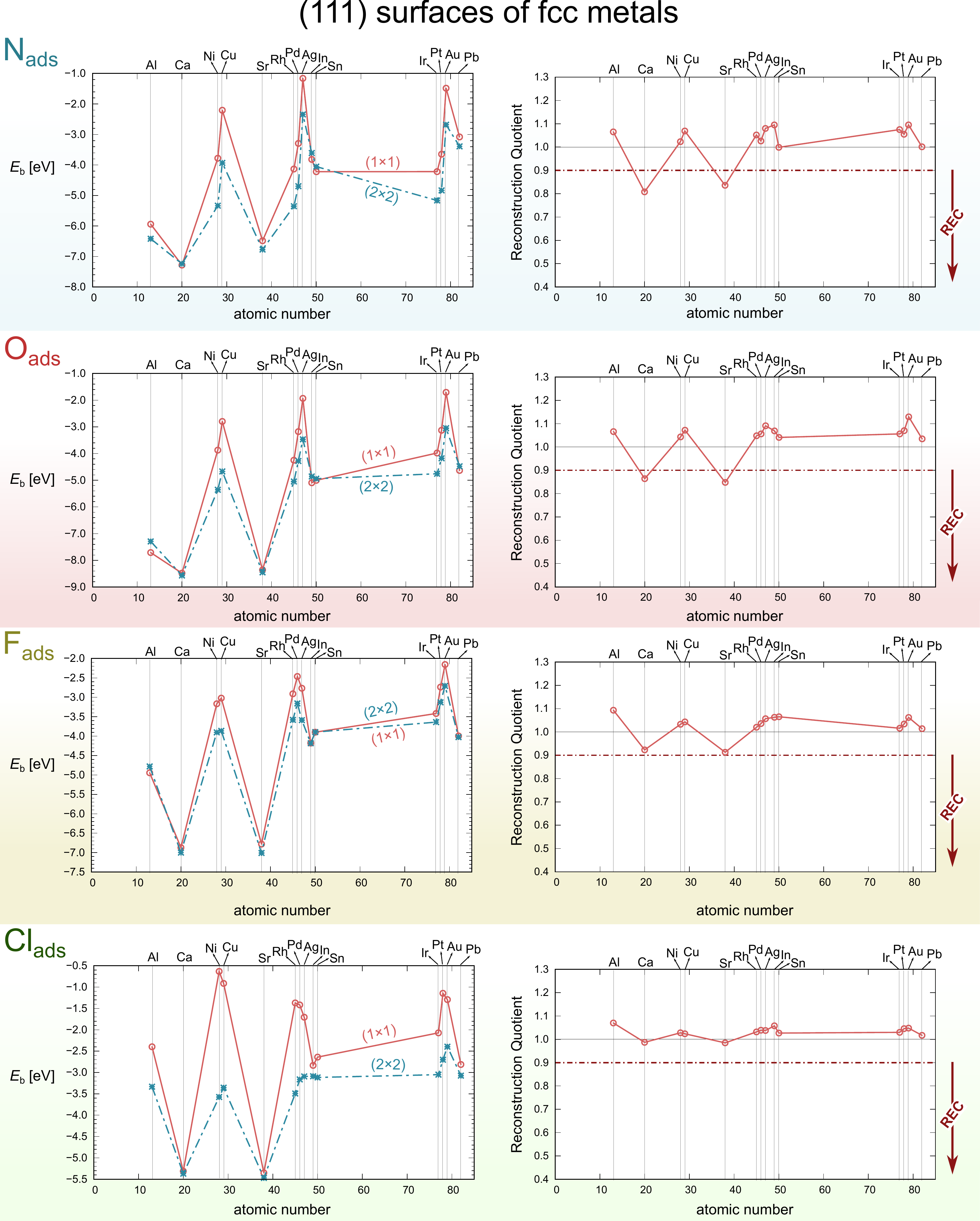}
   \caption{As in \Fig~\ref{fig:Eb-and-ReconDegree-bcc-100}, but
       for fcc(111) surfaces.}
   \label{fig:Eb-and-ReconDegree-fcc-111}
\end{figure*}
\begin{figure*}[htp]
   \centering
   \includegraphics[width=0.97\textwidth]{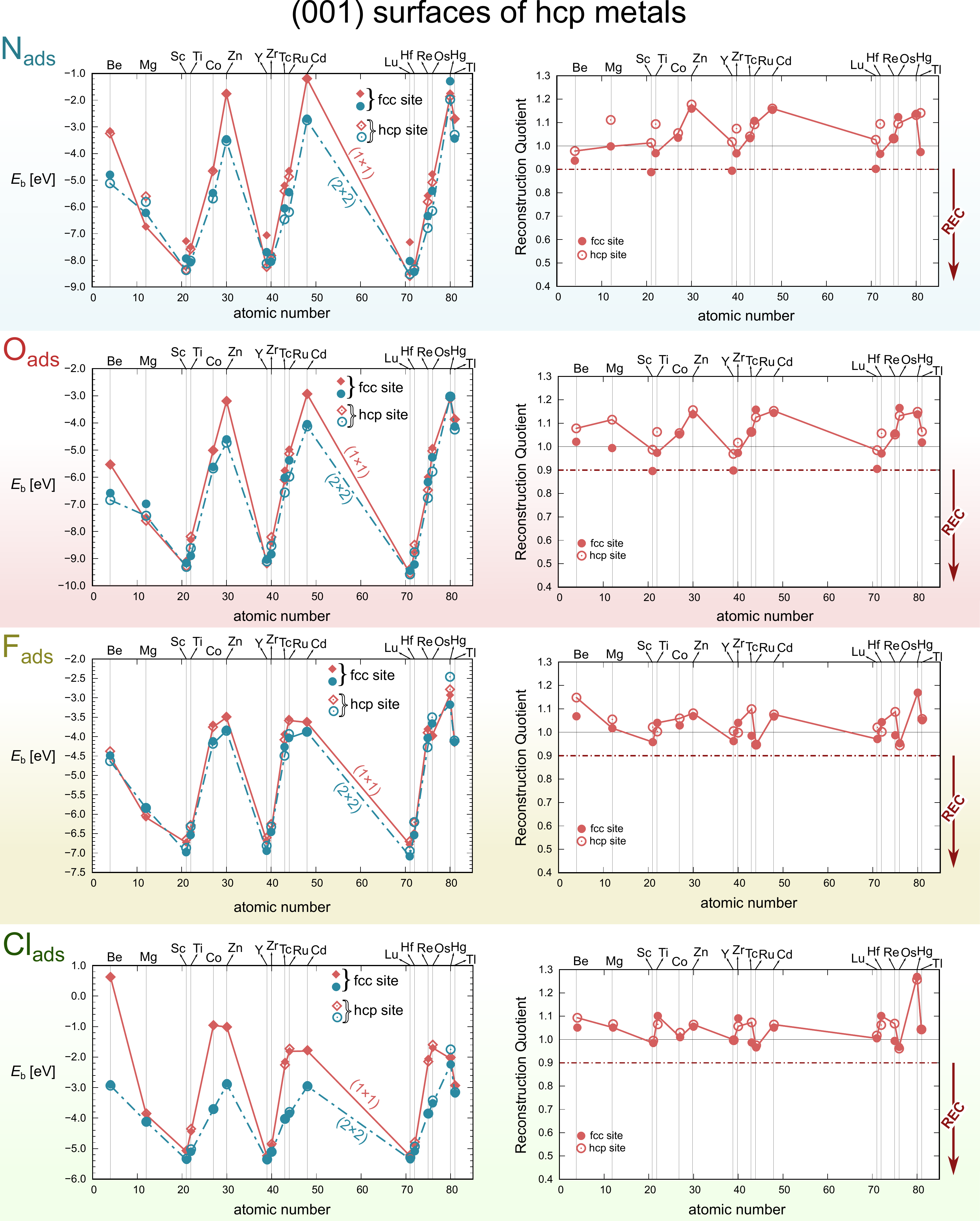}
   \caption{As in \Fig~\ref{fig:Eb-and-ReconDegree-bcc-100}, but
       for hcp(001) surfaces. Filled and empty symbols designate the
     fcc and hcp sites, respectively. The lines connect the most
     stable of the two considered sites and are drawn to guide the
       eye.}
   \label{fig:Eb-and-ReconDegree-hcp-001}
 \end{figure*}

The obtained values for the reconstruction quotient are plotted on the
right-hand side of \Figs~\ref{fig:Eb-and-ReconDegree-bcc-100} to
\ref{fig:Eb-and-ReconDegree-hcp-001}. The plots clearly show that the
extent of reconstruction is the greatest for N and O adatoms on
bcc(100) surfaces, where alkali metals stand out the most. In the case
of fcc metals reconstruction occurs for N and O on the surfaces of Ca
and Sr. According to the \RQ\ value reconstruction also occurs for N
on Sn(100), however, since the \obo\ overlayer is still more stable
this case is labeled as ``attraction'' in the main text.

\subsection{Unoccupied surface area}

\begin{figure}[h]
   \centering
   \includegraphics[width=1.0\columnwidth]{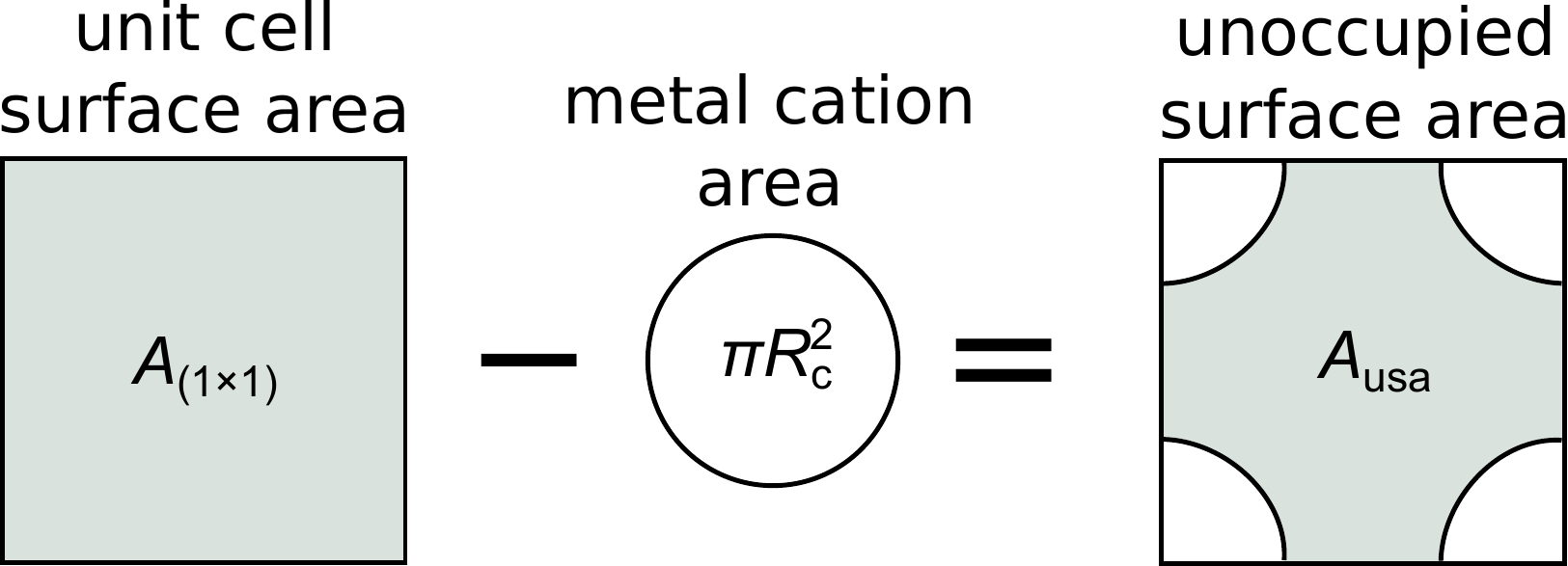}
   \caption[Scheme of unoccupied surface area]{The {\it unoccupied}
     surface area is calculated by subtracting the area occupied by
     the metal cation from the area of the \obo\ unit-cell. For the
     radius of the metal cation, \Rc, the average value of cationic
     radii for all coordination numbers of the lowest cationic
     oxidation state is used. Radii were taken from Shannon and
     Prewitt.\cite{Shannon_ACB26}}
   \label{fig:VacantArea}
\end{figure}

As an approximate {\it ad hoc} criterion for which adatom/metal
  pairs reconstruction can be expected, we defined a quantity termed
  {\it unoccupied surface area}, whose calculation is schematically
  illustrated in \Fig~\ref{fig:VacantArea}.

\section{Stability of adsorption sites}

\begin{figure*}[htp]
  \centering
  \includegraphics[width=0.7\textwidth]{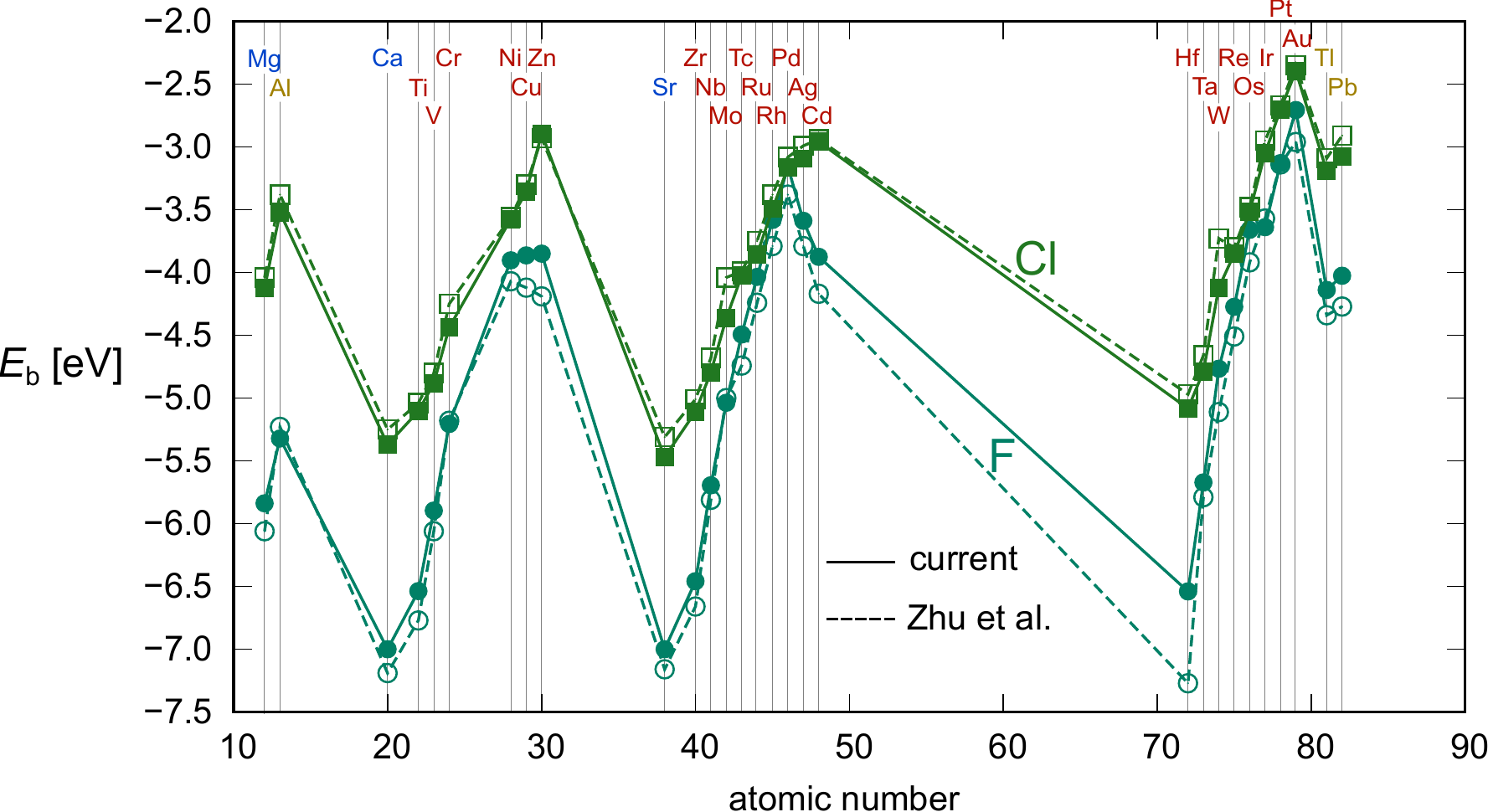}
  \caption{\Eb\ values for \tbt\ overlayers of F and Cl adatoms on
    close-packed metal surfaces obtained in the present study (solid
    lines and filled symbols) as compared to \Eb\ values for F and Cl
    reported by Zhu et al.\cite{Zhu_JESC163} (dashed lines and open
    symbols). Lines are drawn to guide the eye.}
  \label{fig:Eb-current-v-Zhu}
\end{figure*}

The adatoms were mainly adsorbed in hollow sites; for fcc(111)
predominantly fcc-hollow and for hcp(001) both fcc- and hcp-hollow
sites were considered. However, for a few specific cases hollow sites
are either unstable or less stable than bridge or top sites. The cases
where non-hollow sites (or also hcp sites for fcc(111)) were found to
be the stablest are listed in \Tab~\ref{tab:exceptions}. We begin our
analysis by noticing that our results are in line with those reported
by Zhu et al.\cite{Zhu_JESC163} for halogen adatoms (see the
comparison in \Fig~\ref{fig:Eb-current-v-Zhu}). Namely, both F and Cl
prefer the top site on Al(111) at the lower coverage. The top site is
also the preferred site for F on Ir(111) and Pt(111), whereas for Cl
on Ir(111) both fcc and top sites display similar
stabilities. Additionally, on Ca(111) and Sr(111) the hcp site is
found to be more stable than the fcc site for O, F, and Cl,
irrespective of the coverage. Note that most of these exceptions have
been reported by other authors as well.\cite{Roman_PCCP16} As for bcc
metals the top site is preferred for F on W(110). However, we find
that F prefers the hollow site on Mo(110) and not the top site as
reported by Zhu et al.\cite{Zhu_JESC163} The site preference for F
and Cl on hcp metals is also reproduced, the only difference is that
according to our calculations Cl prefers the fcc site on Tc(001),
whereas Zhu et al.\@{} reported that the hcp site is more
stable. Both sets of calculations, however, show that the two sites
have very similar stabilities. The calculated \Eb\ values for \tbt\
layers of F and Cl are compared to those reported by Zhu et al.\@{} in
\Fig~\ref{fig:Eb-current-v-Zhu}. For Cl the average difference between
the two sets of \Eb\ values is $-0.10\pm0.09$~eV, whereas for F the
average difference is $0.19\pm0.15$~eV.

\begin{figure}[t]
  \centering
  \includegraphics[width=1.0\columnwidth]{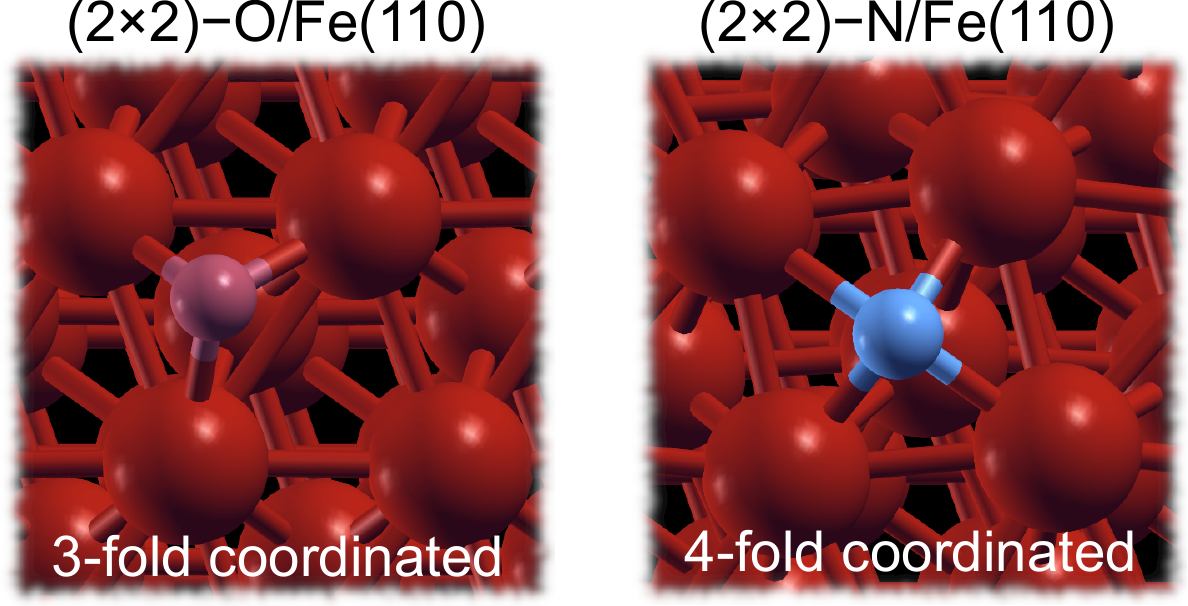}
  \caption{Three- and four-fold hollow sites on the bcc(110)
    surfaces. As an example, the \tbt\ overlayer of O on Fe(110)
    displays three-fold, whereas the \tbt\ overlayer of N on Fe(110)
    displays four-fold coordination.}
  \label{fig:bcc-110-sites}
\end{figure}                                                                

In addition to the already documented site anomalies, we found that F
prefers to adsorb on the top site of Os(001) and bridge site of
Ru(001), whereas the top site is the most stable for Cl on Fe(110),
for Cl on Mn(110) at high coverage, and for F on Bi(100). For O and F
on Al(100) and Fe(100) the bridge site is also the more stable site at
low coverage, whereas at high coverage the two sites generally display
similar stabilities. For Cl on Al(100) and Fe(100) the bridge site is
favored for both investigated coverages.

Finally, as mentioned above adatoms adsorbed on hollow sites of
bcc(110) surfaces display two specific configurations, one where the
adatom is three-fold coordinated and a second one, where the adatom is
four-fold coordinated. Adatoms adsorb predominantly in the three-fold
hollow site, the exceptions are N on Li, Na, K, V, Cr, Fe; O on Li and
Na; and Cl on Li, Na, and Fe. An example of three-fold and four-fold
hollow sites on bcc(110) is shown in \Fig~\ref{fig:bcc-110-sites}.

\section{Adsorption induced work function changes}

\begin{figure*}[ht]
  \centering
  \includegraphics[width=0.7\textwidth]{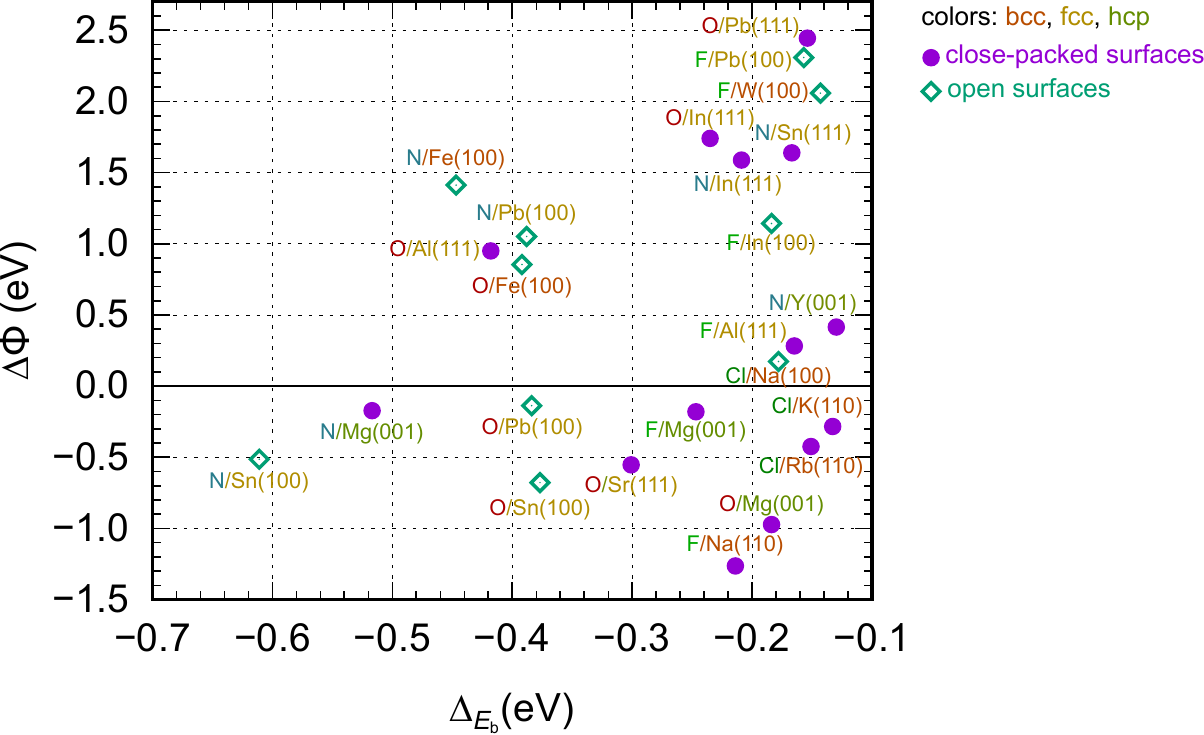}
  \caption{Adsorption induced work function change ($\Delta\Phi$) of
    \obo\ adatom overlayers for all identified cases of attractive
    lateral interactions ($\Del < -0.1$~eV), whereas adsorption
    induced work function changes of all considered adatom overlayers
    are shown in \Fig~\ref{fig:dWf-all}.}
  \label{fig:dWf-vs-dEb}
\end{figure*}

In some cases, such as N, O, and F adatoms on Mg(0001) the lateral
attractive interaction between adatoms are accompanied by an
adsorption induced decrease of the work function.\cite{Cheng_PRL113}
It should be noted, however, that decrease of the work function is not
required for attraction to emerge, as evidenced by
\Fig~\ref{fig:dWf-vs-dEb}, which shows the adsorption induced work
function change for all the ``attraction cases'' currently identified.
Among 24 such identified cases, work function reduces for only 10 of
them.

\begin{figure*}[ht]
  \centering
  \includegraphics[width=1.0\textwidth]{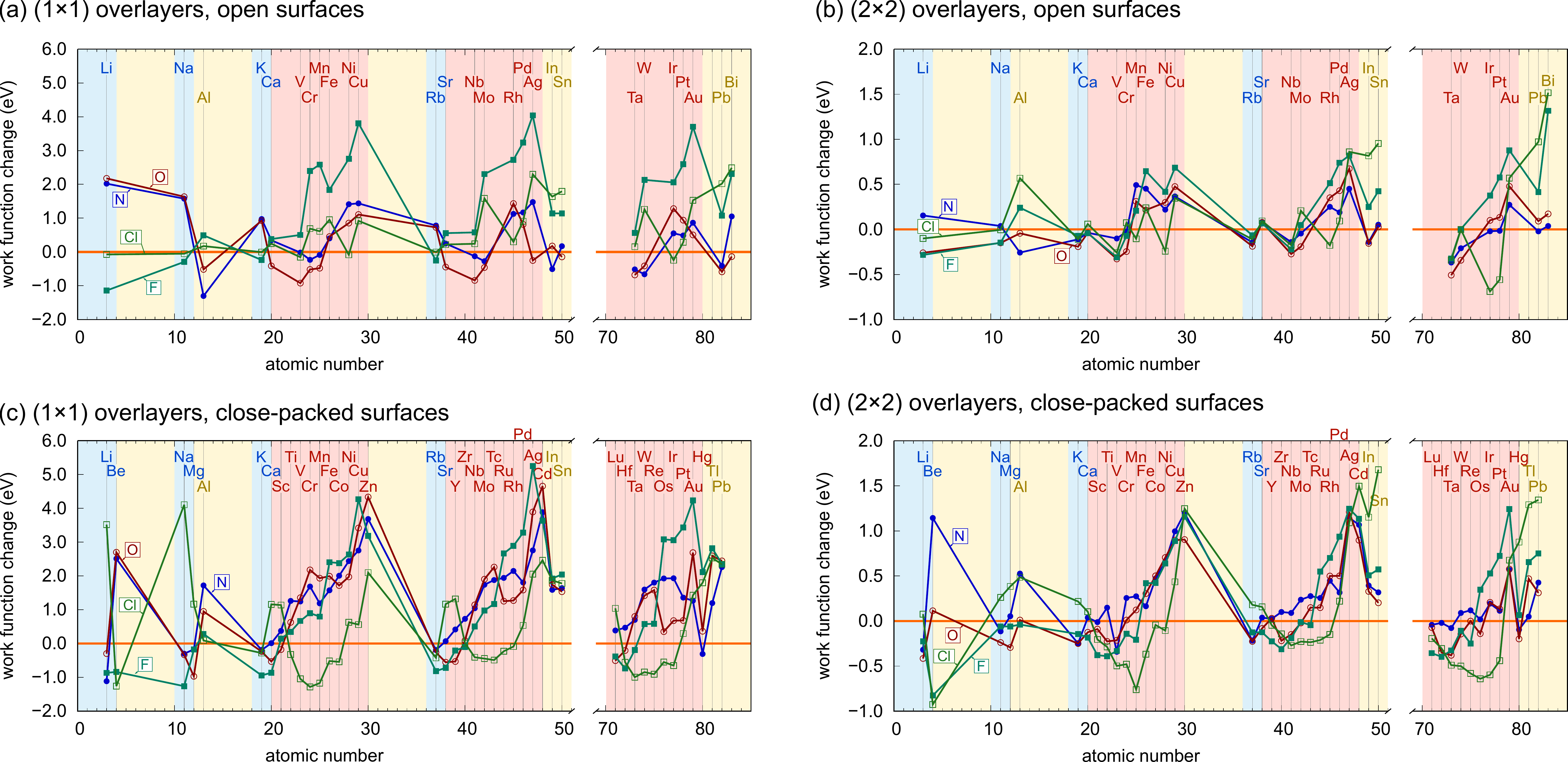}
  \caption{Adsorption induced work function change for all considered
    \obo\ and \tbt\ adatom overlayers. The lines are drawn to guide
    the eye.}
  \label{fig:dWf-all}
\end{figure*}

\begin{figure*}[htp]
   \centering
   \includegraphics[width=0.7\textwidth]{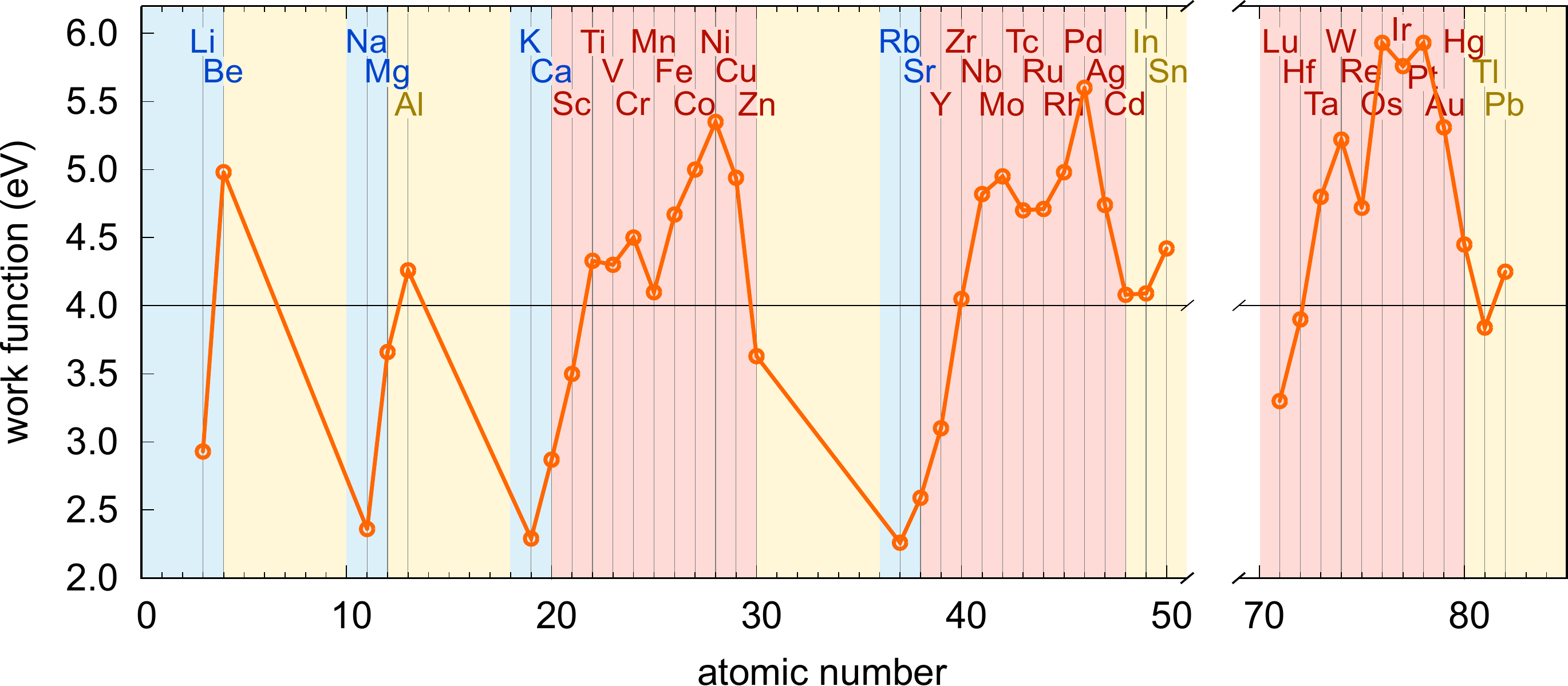}
   \caption{Experimental work functions of metals considered in this
     study; data are taken from Ref.~\citenum{ChemPhysHandbook}.
     Where available the work function for the close-packed surface is
     plotted, otherwise the polycrystalline value is used. The lines
     are drawn to guide the eye.}
   \label{fig:work-function}
\end{figure*}

\Fig~\ref{fig:dWf-all} plots the adsorption induced work function
changes for all the considered adatom overlayers and
\Fig~\ref{fig:work-function} shows the experimental work functions for
either closed-packed or polycrystalline surfaces of the 44 metals
considered in this study.


\bibliography{biblio,footnotes}

%
\begin{table*}[htb]
  \centering
  \caption{Binding energies for \obo\ and \tbt\ overlayers of the four
    considered adatoms adsorbed on hollow sites of the bcc(100) metal
      surfaces. All values are in eV/adatom.}
  \begin{threeparttable}
  
\begin{tabular}{l@{\hskip 0.4cm}cc@{\hskip 0.4cm}cc@{\hskip 0.4cm}cc@{\hskip 0.4cm}cc@{\hskip 0.4cm}}
  \hline
  ~\\[-0.98em]
Surface & \multicolumn{2}{@{\hskip -0.2cm}c}{$\mathrm{N_{ads}}$  } &  \multicolumn{2}{@{\hskip -0.2cm}c}{$\mathrm{O_{ads}}$ } &  \multicolumn{2}{@{\hskip -0.2cm}c}{$\mathrm{F_{ads}}$ } &  \multicolumn{2}{@{\hskip -0.2cm}c}{$\mathrm{Cl_{ads}}$ } \\
  & \tbt\ & \obo\  & \tbt\ & \obo\ & \tbt\ & \obo\ & \tbt\ & \obo\   \\
\hline
Li(100) & $ -6.25 $ & $ -3.37 $ & $ -8.27 $ & $ -5.89 $ & $ -6.52 $ & $ -6.36 $ & $ -5.20 $ & $ -5.26 $ \\  
Na(100) & $ -3.96 $ & $ -1.93 $ & $ -6.24 $ & $ -4.38 $ & $ -5.90 $ & $ -5.62 $ & $ -4.86 $ & $ -5.03 $ \\  
K(100) & $ -3.54 $ & $ -1.32 $ & $ -6.07 $ & $ -3.85 $ & $ -5.96 $ & $ -5.43 $ & $ -5.15 $ & $ -5.05 $ \\  
V(100) & $ -8.48 $ & $ -8.11 $ & $ -8.16 $ & $ -7.81 $ & $ -5.18 $ & $ -4.87 $ & $ -4.58 $ & $ -3.84 $ \\  
Cr(100) & $ -8.23 $ & $ -7.44 $ & $ -7.59 $ & $ -6.71 $ & $ -4.56 $ & $ -4.39 $ & $ -4.75 $ & $ -3.20 $ \\  
Mn(100) & $ -7.57 $ & $ -6.74 $ & $ -7.22 $ & $ -5.72 $ & $ -4.99 $ & $ -3.97 $ & $ -4.87 $ & $ -2.67 $ \\  
Fe(100) & $ -5.82 $ & $ -6.27 $ & $ -5.25 $\tnote{a} & $ -5.64 $ & $ -5.31 $\tnote{a} & $ -4.46 $\tnote{a} & $ -3.42 $\tnote{a} & $ -3.20 $\tnote{a} \\  
Rb(100) & $ -3.36 $ & $ -1.17 $ & $ -5.90 $ & $ -3.70 $ & $ -5.97 $ & $ -5.34 $ & $ -5.14 $ & $ -4.97 $ \\  
Nb(100) & $ -8.24 $ & $ -7.75 $ & $ -7.75 $ & $ -7.31 $ & $ -5.00 $ & $ -4.77 $ & $ -4.51 $ & $ -4.09 $ \\  
Mo(100) & $ -7.56 $ & $ -6.99 $ & $ -6.83 $ & $ -6.02 $ & $ -4.27 $ & $ -4.15 $ & $ -4.60 $ & $ -3.51 $ \\  
Ta(100) & $ -8.32 $ & $ -7.72 $ & $ -7.64 $ & $ -6.99 $ & $ -4.76 $ & $ -4.54 $ & $ -4.31 $ & $ -3.85 $ \\  
W(100) & $ -7.73 $ & $ -6.74 $ & $ -6.66 $ & $ -5.58 $ & $ -3.88 $ & $ -4.03 $ & $ -4.47 $ & $ -3.44 $ \\  
\hline
\end{tabular}
\begin{tablenotes}
\item{$^{\mathrm a}$ The reported value is for the more stable bridge site.}
\end{tablenotes}
\end{threeparttable}
  \label{tab:Eb-bcc-100}
\end{table*}

\begin{table*}[htb]
  \centering
  \caption{As in \Tab~\ref{tab:Eb-bcc-100}, but for hollow sites
      on the bcc(110) surfaces.
      All values are in eV/adatom.}
  \begin{threeparttable}
  
\begin{tabular}{l@{\hskip 0.4cm}cc@{\hskip 0.4cm}cc@{\hskip 0.4cm}cc@{\hskip 0.4cm}cc@{\hskip 0.4cm}}
  \hline
  ~\\[-0.98em]
Surface & \multicolumn{2}{@{\hskip -0.2cm}c}{$\mathrm{N_{ads}}$  } &  \multicolumn{2}{@{\hskip -0.2cm}c}{$\mathrm{O_{ads}}$ } &  \multicolumn{2}{@{\hskip -0.2cm}c}{$\mathrm{F_{ads}}$ } &  \multicolumn{2}{@{\hskip -0.2cm}c}{$\mathrm{Cl_{ads}}$ } \\
  & \tbt\ & \obo\  & \tbt\ & \obo\ & \tbt\ & \obo\ & \tbt\ & \obo\   \\
\hline
Li(110) & $ -6.71 $ & $ -5.61 $ & $ -8.44 $ & $ -7.84 $ & $ -6.76 $ & $ -6.77 $ & $ -5.21 $ & $ -3.75 $ \\  
Na(110) & $ -3.94 $ & $ -3.26 $ & $ -6.32 $ & $ -6.07 $ & $ -5.96 $ & $ -6.17 $ & $ -4.83 $ & $ -4.15 $ \\  
K(110) & $ -3.52 $ & $ -2.62 $ & $ -6.00 $ & $ -5.42 $ & $ -6.13 $ & $ -6.12 $ & $ -5.03 $ & $ -5.16 $ \\  
V(110) & $ -7.82 $ & $ -6.41 $ & $ -8.05 $ & $ -7.15 $ & $ -5.90 $ & $ -5.71 $ & $ -4.88 $ & $ -2.90 $ \\  
Cr(110) & $ -7.25 $ & $ -5.55 $ & $ -7.47 $ & $ -6.45 $ & $ -5.21 $ & $ -4.76 $ & $ -4.44 $ & $ -1.52 $ \\  
Mn(110) & $ -6.95 $ & $ -5.38 $ & $ -7.17 $ & $ -6.02 $ & $ -4.98 $ & $ -4.13 $ & $ -4.53 $ & $ -1.22 $\tnote{a} \\  
Fe(110) & $ -6.27 $ & $ -5.06 $ & $ -5.96 $ & $ -5.62 $ & $ -4.01 $ & $ -4.00 $ & $ -3.76 $\tnote{a} & $ -1.57 $\tnote{a} \\  
Rb(110) & $ -3.09 $ & $ -2.40 $ & $ -5.96 $ & $ -5.14 $ & $ -6.08 $ & $ -6.03 $ & $ -5.02 $ & $ -5.17 $ \\  
Nb(110) & $ -7.19 $ & $ -6.66 $ & $ -7.69 $ & $ -7.20 $ & $ -5.70 $ & $ -5.67 $ & $ -4.79 $ & $ -3.90 $ \\  
Mo(110) & $ -7.00 $ & $ -5.89 $ & $ -7.21 $ & $ -6.67 $ & $ -5.04 $ & $ -4.79 $ & $ -4.36 $ & $ -2.85 $ \\  
Ta(110) & $ -7.68 $ & $ -7.10 $ & $ -8.02 $ & $ -7.54 $ & $ -5.67 $ & $ -5.68 $ & $ -4.79 $ & $ -3.95 $ \\  
W(110) & $ -7.15 $ & $ -6.13 $ & $ -7.19 $ & $ -6.87 $ & $ -5.15 $\tnote{a} & $ -4.87 $\tnote{a} & $ -4.12 $ & $ -2.78 $ \\  
\hline
\end{tabular}
  \begin{tablenotes}
  \item{$^\mathrm{a}$} The reported value is for the more stable top site.
  \end{tablenotes} 
\end{threeparttable}

  \label{tab:Eb-bcc-110}
\end{table*}

\begin{table*}[htb]
  \centering
  \caption{As in \Tab~\ref{tab:Eb-bcc-100}, but for fcc sites on
      the fcc(111) surfaces.
    All values are in eV/adatom.}
  \begin{threeparttable}    
  \begin{tabular}{l@{\hskip 0.4cm}cc@{\hskip 0.4cm}cc@{\hskip 0.4cm}cc@{\hskip 0.4cm}cc@{\hskip 0.4cm}}
  \hline
  ~\\[-0.98em]
Surface & \multicolumn{2}{@{\hskip -0.2cm}c}{$\mathrm{N_{ads}}$  } &  \multicolumn{2}{@{\hskip -0.2cm}c}{$\mathrm{O_{ads}}$ } &  \multicolumn{2}{@{\hskip -0.2cm}c}{$\mathrm{F_{ads}}$ } &  \multicolumn{2}{@{\hskip -0.2cm}c}{$\mathrm{Cl_{ads}}$ } \\
  & \tbt\ & \obo\  & \tbt\ & \obo\ & \tbt\ & \obo\ & \tbt\ & \obo\   \\
\hline
Al(111) & $ -6.42 $ & $ -5.94 $ & $ -7.29 $ & $ -7.71 $ & $ -5.05 $\tnote{a} & $ -4.95 $ & $ -3.52 $\tnote{a} & $ -2.40 $ \\  
Ca(111) & $ -7.24 $ & $ -7.28 $ & $ -8.57 $\tnote{b} & $ -8.48 $\tnote{b} & $ -7.00 $\tnote{b} & $ -6.87 $\tnote{b} & $ -5.37 $\tnote{b} & $ -5.31 $\tnote{b} \\  
Ni(111) & $ -5.33 $ & $ -3.78 $ & $ -5.36 $ & $ -3.87 $ & $ -3.90 $ & $ -3.17 $ & $ -3.57 $ & $ -0.63 $ \\  
Cu(111) & $ -3.93 $ & $ -2.20 $ & $ -4.67 $ & $ -2.80 $ & $ -3.86 $ & $ -3.02 $ & $ -3.36 $ & $ -0.91 $ \\  
Sr(111) & $ -6.77 $\tnote{b} & $ -6.48 $ & $ -8.45 $\tnote{b} & $ -8.36 $\tnote{b} & $ -7.00 $\tnote{b} & $ -6.78 $\tnote{b} & $ -5.47 $\tnote{b} & $ -5.35 $\tnote{b} \\  
Rh(111) & $ -5.35 $ & $ -4.13 $ & $ -5.05 $ & $ -4.25 $ & $ -3.58 $ & $ -2.91 $ & $ -3.49 $ & $ -1.37 $ \\  
Pd(111) & $ -4.69 $ & $ -3.29 $ & $ -4.28 $ & $ -3.18 $ & $ -3.16 $ & $ -2.46 $ & $ -3.16 $ & $ -1.42 $ \\  
Ag(111) & $ -2.34 $ & $ -1.16 $ & $ -3.47 $ & $ -1.93 $ & $ -3.59 $ & $ -2.77 $ & $ -3.09 $ & $ -1.71 $ \\  
In(111) & $ -3.61 $ & $ -3.82 $ & $ -4.86 $ & $ -5.09 $ & $ -4.18 $ & $ -4.18 $ & $ -3.09 $ & $ -2.83 $ \\  
Sn(111) & $ -4.05 $ & $ -4.22 $ & $ -4.94 $ & $ -5.00 $ & $ -3.90 $ & $ -3.90 $ & $ -3.12 $ & $ -2.64 $ \\  
Ir(111) & $ -5.16 $ & $ -4.22 $ & $ -4.76 $ & $ -3.98 $ & $ -3.46 $\tnote{a} & $ -3.42$\tnote{a}& $ -3.05 $\tnote{a} & $ -2.07 $ \\  
Pt(111) & $ -4.83 $ & $ -3.65 $ & $ -4.18 $ & $ -3.13 $ & $ -3.13 $\tnote{a} & $ -2.74\tnote{a} $ & $ -2.70 $ & $ -1.15 $ \\  
Au(111) & $ -2.68 $ & $ -1.48 $ & $ -3.05 $ & $ -1.71 $ & $ -2.71 $ & $ -2.15 $ & $ -2.40 $ & $ -1.29 $ \\  
Pb(111) & $ -3.39 $ & $ -3.08 $ & $ -4.48 $ & $ -4.63 $ & $ -4.03 $ & $ -4.00 $ & $ -3.07 $ & $ -2.81 $ \\  
\hline
  \end{tabular}
  \begin{tablenotes}
  \item{$^\mathrm{a}$} The reported value is for the more stable top site.
  \item{$^\mathrm{b}$} The reported value is for the more stable hcp site.
 \end{tablenotes} 
\end{threeparttable}

  \label{tab:Eb-fcc-111}
\end{table*}

\begin{table*}[htb]
  \centering
  \caption{As in \Tab~\ref{tab:Eb-bcc-100}, but for hollow sites
      on the fcc(100)
      surfaces.
    All values are in eV/adatom.}
  \begin{threeparttable}
  
\begin{tabular}{l@{\hskip 0.4cm}cc@{\hskip 0.4cm}cc@{\hskip 0.4cm}cc@{\hskip 0.4cm}cc@{\hskip 0.4cm}}
  \hline
  ~\\[-0.98em]
Surface & \multicolumn{2}{@{\hskip -0.2cm}c}{$\mathrm{N_{ads}}$  } &  \multicolumn{2}{@{\hskip -0.2cm}c}{$\mathrm{O_{ads}}$ } &  \multicolumn{2}{@{\hskip -0.2cm}c}{$\mathrm{F_{ads}}$ } &  \multicolumn{2}{@{\hskip -0.2cm}c}{$\mathrm{Cl_{ads}}$ } \\
  & \tbt\ & \obo\  & \tbt\ & \obo\ & \tbt\ & \obo\ & \tbt\ & \obo\   \\
\hline
Al(100) & $ -5.77 $ & $ -6.69 $ & $ -6.60 $\tnote{a} & $ -6.65 $ & $ -5.23 $\tnote{a} & $ -5.29 $ & $ -3.55 $\tnote{a} & $ -2.86 $\tnote{a} \\  
Ca(100) & $ -7.36 $ & $ -6.12 $ & $ -8.19 $ & $ -8.11 $ & $ -6.16 $ & $ -6.02 $ & $ -5.12 $ & $ -5.12 $ \\  
Ni(100) & $ -6.20 $ & $ -4.20 $ & $ -5.61 $ & $ -4.13 $ & $ -3.96 $ & $ -3.36 $ & $ -3.83 $ & $ -1.12 $ \\  
Cu(100) & $ -4.66 $ & $ -3.07 $ & $ -4.94 $ & $ -3.61 $ & $ -3.95 $ & $ -3.35 $ & $ -3.50 $ & $ -1.33 $ \\  
Sr(100) & $ -6.69 $ & $ -5.45 $ & $ -7.80 $ & $ -7.24 $ & $ -6.17 $ & $ -6.01 $ & $ -5.19 $ & $ -5.16 $ \\  
Rh(100) & $ -5.69 $ & $ -4.26 $ & $ -5.04 $ & $ -3.84 $ & $ -3.59 $ & $ -3.10 $ & $ -3.71 $ & $ -1.80 $ \\  
Pd(100) & $ -4.86 $ & $ -3.29 $ & $ -4.24 $ & $ -3.26 $ & $ -3.25 $ & $ -2.67 $ & $ -3.39 $ & $ -1.79 $ \\  
Ag(100) & $ -2.77 $ & $ -2.10 $ & $ -3.70 $ & $ -2.98 $ & $ -3.69 $ & $ -3.19 $ & $ -3.20 $ & $ -1.98 $ \\  
In(100) & $ -3.77 $ & $ -3.84 $ & $ -4.85 $ & $ -4.64 $ & $ -4.09 $ & $ -4.28 $ & $ -3.11 $ & $ -3.08 $ \\  
Sn(100) & $ -3.73 $ & $ -4.34 $ & $ -4.32 $ & $ -4.69 $ & $ -3.79 $ & $ -3.89 $ & $ -2.85 $ & $ -2.75 $ \\  
Ir(100) & $ -5.41 $ & $ -4.03 $ & $ -4.66 $ & $ -3.62 $ & $ -3.04 $ & $ -2.83 $ & $ -3.21 $ & $ -1.56 $ \\  
Pt(100) & $ -4.65 $ & $ -3.17 $ & $ -3.87 $ & $ -2.78 $ & $ -2.65 $ & $ -2.36 $ & $ -2.85 $ & $ -1.50 $ \\  
Au(100) & $ -2.47 $ & $ -2.04 $ & $ -2.84 $ & $ -2.18 $ & $ -2.77 $ & $ -2.42 $ & $ -2.46 $ & $ -1.51 $ \\  
Pb(100) & $ -3.34 $ & $ -3.73 $ & $ -4.25 $ & $ -4.63 $ & $ -3.98 $ & $ -4.13 $ & $ -3.09 $ & $ -2.97 $ \\  
\hline
\end{tabular}
\begin{tablenotes}
 \item{$^{\mathrm{a}}$ The reported value is for the more stable bridge site.} 
\end{tablenotes}
\end{threeparttable}

  \label{tab:Eb-fcc-100}
\end{table*}
%

\begin{table*}[htb]
  \centering
  \caption{As in \Tab~\ref{tab:Eb-bcc-100}, but for fcc and hcp
      sites on the hcp(001) surfaces. The most stable site is emphasized
      in bold for each particular case.
      All values are in eV/adatom.}
    \scriptsize
    {\onehalfspacing
      \begin{threeparttable}
  \begin{tabular}{l@{\hskip 0.4cm}cc@{\hskip 0.4cm}cc@{\hskip 0.4cm}cc@{\hskip 0.4cm}cc@{\hskip 0.4cm}}
  \hline
  ~\\[-0.98em]
Surface & \multicolumn{2}{@{\hskip -0.2cm}c}{$\mathrm{N_{ads}}$  } &  \multicolumn{2}{@{\hskip -0.2cm}c}{$\mathrm{O_{ads}}$ } &  \multicolumn{2}{@{\hskip -0.2cm}c}{$\mathrm{F_{ads}}$ } &  \multicolumn{2}{@{\hskip -0.2cm}c}{$\mathrm{Cl_{ads}}$ } \\
  & \tbt\ & \obo\  & \tbt\ & \obo\ & \tbt\ & \obo\ & \tbt\ & \obo\   \\
  & fcc/hcp & fcc/hcp  & fcc/hcp & fcc/hcp & fcc/hcp & fcc/hcp & fcc/hcp & fcc/hcp   \\
\hline
Be(001) & $-4.79$/$\mathbf{-5.12}$ & $-3.16$/$\mathbf{-3.25}$ & $-6.59$/$\mathbf{-6.84}$ & $-5.51$/$\mathbf{-5.53}$ & $-4.49$/$\mathbf{-4.63}$  & $\mathbf{-4.51}$/$-4.38$ & $-2.89$/$\mathbf{-2.94}$  & $\mathbf{+0.63}$/$+0.63$ \\
Mg(001)  & $\mathbf{-6.22}$/$-5.81$  & $\mathbf{-6.74}$/$-5.60$ & $-6.98$/$\mathbf{-7.42}$ & $-7.48$/$\mathbf{-7.60}$  & $-5.84$/$-5.84$  & $\mathbf{-6.08}$/$-6.04$  & $\mathbf{-4.12}$/$-4.12$  & $\mathbf{-3.89}$/$-3.85$ \\
Sc(001) & $-7.93$/$\mathbf{-8.37}$ & $-7.28$/$\mathbf{-8.37}$ & $-9.16$/$\mathbf{-9.30}$ & $-9.07$/$\mathbf{-9.29}$  & $\mathbf{-6.98}$/$-6.87$  & $\mathbf{-6.73}$/$-6.66$  & $\mathbf{-5.37}$/$-5.33$  & $\mathbf{-5.10}$/$-5.07$ \\
Ti(001)  & $\mathbf{-8.07}$/$-8.01$ & $-7.50$/$\mathbf{-7.58}$  & $\mathbf{-8.90}$/$-8.62$  & $\mathbf{-8.31}$/$-8.19$  & $\mathbf{-6.54}$/$-6.33$  & $\mathbf{-6.46}$/$-6.28$  & $\mathbf{-5.10}$/$-5.02$  & $\mathbf{-4.44}$/$-4.35$ \\
Co(001) & $-5.48$/$\mathbf{-5.68}$ & $-4.61$/$\mathbf{-4.66}$ & $-5.62$/$\mathbf{-5.68}$ & $-5.00$/$\mathbf{-5.01}$ & $-4.12$/$\mathbf{-4.19}$ & $-3.69$/$\mathbf{-3.75}$  & $\mathbf{-3.71}$/$-3.71$  & $\mathbf{-0.96}$/$-0.96$ \\
Zn(001) & $-3.46$/$\mathbf{-3.53}$ & $-1.73$/$\mathbf{-1.76}$ & $-4.61$/$\mathbf{-4.73}$ & $-3.19$/$\mathbf{-3.19}$ & $-3.82$/$\mathbf{-3.85}$  & $\mathbf{-3.51}$/$-3.49$ & $-2.86$/$\mathbf{-2.89}$  & $\mathbf{-1.02}$/$-1.01$ \\
Y(001) & $-7.69$/$\mathbf{-8.12}$ & $-7.06$/$\mathbf{-8.25}$ & $-9.03$/$\mathbf{-9.10}$ & $-8.99$/$\mathbf{-9.17}$  & $\mathbf{-6.94}$/$-6.82$  & $\mathbf{-6.69}$/$-6.60$  & $\mathbf{-5.38}$/$-5.35$  & $\mathbf{-5.27}$/$-5.24$ \\
Zr(001)  & $\mathbf{-8.05}$/$-7.95$ & $-7.76$/$\mathbf{-7.82}$  & $\mathbf{-8.84}$/$-8.53$  & $\mathbf{-8.41}$/$-8.21$  & $\mathbf{-6.46}$/$-6.32$  & $\mathbf{-6.41}$/$-6.25$ & $-5.09$/$\mathbf{-5.11}$  & $\mathbf{-4.90}$/$-4.84$ \\
Tc(001) & $-6.05$/$\mathbf{-6.46}$ & $-5.19$/$\mathbf{-5.40}$ & $-6.02$/$\mathbf{-6.56}$ & $-5.76$/$\mathbf{-6.09}$ & $-4.27$/$\mathbf{-4.49}$ & $-3.94$/$\mathbf{-4.08}$ & $-4.02$/$\mathbf{-4.03}$ & $-2.16$/$\mathbf{-2.26}$ \\
Ru(001) & $-5.45$/$\mathbf{-6.19}$ & $-4.65$/$\mathbf{-4.87}$ & $-5.38$/$\mathbf{-5.98}$ & $-4.98$/$\mathbf{-5.13}$  & $\mathbf{-4.14}$\tnote{b}/$-3.93$ & $-3.54$/$\mathbf{-3.59}$\tnote{b}  & $\mathbf{-3.85}$/$-3.80$  & $\mathbf{-1.81}$/$-1.72$ \\
Cd(001) & $-2.69$/$\mathbf{-2.75}$ & $-1.18$/$\mathbf{-1.19}$ & $-4.05$/$\mathbf{-4.13}$  & $\mathbf{-2.94}$/$-2.93$ & $-3.86$/$\mathbf{-3.87}$  & $\mathbf{-3.63}$/$-3.62$ & $-2.92$/$\mathbf{-2.96}$  & $\mathbf{-1.80}$/$-1.78$ \\
Lu(001) & $-8.03$/$\mathbf{-8.52}$ & $-7.32$/$\mathbf{-8.59}$ & $-9.44$/$\mathbf{-9.58}$ & $-9.35$/$\mathbf{-9.55}$  & $\mathbf{-7.08}$/$-6.95$  & $\mathbf{-6.78}$/$-6.70$  & $\mathbf{-5.35}$/$-5.29$  & $\mathbf{-5.25}$/$-5.21$ \\
Hf(001)  & $\mathbf{-8.43}$/$-8.34$ & $-8.13$/$\mathbf{-8.28}$  & $\mathbf{-9.22}$/$-8.77$  & $\mathbf{-8.77}$/$-8.49$  & $\mathbf{-6.54}$/$-6.20$  & $\mathbf{-6.49}$/$-6.21$  & $\mathbf{-5.08}$/$-4.93$  & $\mathbf{-4.90}$/$-4.78$ \\
Re(001) & $-6.34$/$\mathbf{-6.79}$ & $-5.58$/$\mathbf{-5.81}$ & $-6.18$/$\mathbf{-6.77}$ & $-6.00$/$\mathbf{-6.47}$ & $-4.04$/$\mathbf{-4.27}$ & $-3.81$/$\mathbf{-3.90}$ & $-3.83$/$\mathbf{-3.85}$ & $-2.06$/$\mathbf{-2.15}$ \\
Os(001) & $-5.40$/$\mathbf{-6.15}$ & $-4.77$/$\mathbf{-5.08}$ & $-5.27$/$\mathbf{-5.79}$ & $-4.92$/$\mathbf{-5.04}$  & $\mathbf{-3.66}$/$-3.50$  & $-3.98$\tnote{a}/$-3.98$\tnote{a}  & $\mathbf{-3.52}$/$-3.42$  & $\mathbf{-1.70}$/$-1.60$ \\
Hg(001) & $-1.29$/$\mathbf{-1.98}$ & $-1.75$/$\mathbf{-1.87}$ & $-3.03$/$\mathbf{-3.03}$ & $-3.06$/$\mathbf{-3.10}$  & $\mathbf{-3.18}$/$-2.46$  & $\mathbf{-2.93}$/$-2.78$  & $\mathbf{-2.24}$/$-1.75$  & $\mathbf{-2.05}$/$-2.01$ \\
Tl(001)  & $\mathbf{-3.44}$/$-3.29$ & $-2.66$/$\mathbf{-2.70}$ & $-4.14$/$\mathbf{-4.24}$  & $\mathbf{-3.87}$/$-3.87$  & $\mathbf{-4.14}$/$-4.10$ & $-4.13$/$\mathbf{-4.14}$  & $\mathbf{-3.19}$/$-3.15$  & $\mathbf{-2.94}$/$-2.93$ \\
\hline
  \end{tabular}
    \begin{tablenotes}
  \item{$^\mathrm{a}$} The reported value is for the more stable top site.
  \item{$^\mathrm{b}$} The reported value is for the more stable bridge site.
  \end{tablenotes}
\end{threeparttable}


    }
  \label{tab:Eb-hcp-001-hcpsite}
\end{table*}

\begin{table*}[htb]
  \centering
  \caption{As in \Tab~\ref{tab:Eb-bcc-100}, but for the hollow
      sites on sc(100) surfaces. Note that only Bi is considered as a
      simple-cubic metal for the reasons explained in the main text.
    All values are in eV/adatom.}
  \begin{threeparttable}
\begin{tabular}{l@{\hskip 0.4cm}cc@{\hskip 0.4cm}cc@{\hskip 0.4cm}cc@{\hskip 0.4cm}cc@{\hskip 0.4cm}}
  \hline
  ~\\[-0.98em]
Surface & \multicolumn{2}{@{\hskip -0.2cm}c}{$\mathrm{N_{ads}}$  } &  \multicolumn{2}{@{\hskip -0.2cm}c}{$\mathrm{O_{ads}}$  } &  \multicolumn{2}{@{\hskip -0.2cm}c}{$\mathrm{F_{ads}}$ } &  \multicolumn{2}{@{\hskip -0.2cm}c}{$\mathrm{Cl_{ads}}$ } \\
& \tbt\ & \obo\  & \tbt\ & \obo\ & \tbt\ & \obo\ & \tbt\ & \obo\   \\
\hline
Bi(100) & $ -2.29 $ & $ -3.13 $ & $ -3.14 $ & $ -4.21 $ & $ -4.16 $\tnote{a} & $ -3.20 $ & $ -2.42 $ & $ -2.07 $ \\  
\hline
\end{tabular}
\begin{tablenotes}
\item{$^\mathrm{a}$} Adatom relaxed to top site (hollow site is unstable)
\end{tablenotes}
\end{threeparttable}


  \label{tab:Eb-sc-100}
\end{table*}

\begin{table*}[htb]
  \centering
  \caption{List of exceptions, where the adatoms do not
      preferentially adsorb into ``default'' hollow sites; for fcc(111)
      the fcc site is considered as the default hollow site.
    \Eb\ values of \obo\ and \tbt\ overlayers of adatoms adsorbed
      on the default hollow site are
    compared to those on the alternative site, which is the most
      stable for at least one considered overlayer. All values are in
    eV/adatom.}      
\begin{tabular}{l@{\hskip 0.4cm}ccc@{\hskip 0.4cm}ccc}
  \hline
  ~\\[-0.98em]
adsorbate/surface & default hollow site & \multicolumn{2}{@{\hskip -0.2cm}c}{\Eb} & alternative site &  \multicolumn{2}{c}{\Eb}\\
system  & & \tbt\ & \obo\ & & \tbt\ & \obo\   \\
\hline
  F/Al(111) & fcc &    $-4.78$     & $-4.95$    & top  & $-5.05$ & $-4.59$ \\
  Cl/Al(111) & fcc &   $-3.33$     & $-2.40$    & top  & $-3.52$ & $-2.20$ \\
  O/Al(100) & hollow  & $-5.96$   &    $-6.65$   & bridge & $-6.60$ &  $-6.53$ \\ 
  F/Al(100) & hollow  & $-4.22$     &  $-5.29$  & bridge & $-5.23$  &   $-5.29$ \\
  Cl/Al(100) & hollow & $-3.10$     & $-2.53$   & bridge & $-3.55$   &  $-2.86$ \\
  O/Ca(111) & fcc &    $-8.36$     & $-7.93$    & hcp  & $-8.57$  & $ -8.48$   \\
  F/Ca(111) & fcc &    $-6.86$     & $-6.80$    & hcp  & $-7.00$  & $-6.87$ \\
  Cl/Ca(111) & fcc &   $-5.29$    & $-5.27$     & hcp  & $-5.37$  & $-5.31$ \\
  Cl/Mn(110)& hollow  &  $-5.80$     & /   & top  &   /       & $-1.43$  \\
  Cl/Fe(110)& hollow  &  $-3.33$     & /   & top  &  $-3.76$  & $-1.57$  \\
  O/Fe(100) & hollow &   $-5.08$     &   $-5.64$  & bridge & $-5.25$ & $-5.64$  \\ 
  F/Fe(100) & hollow  &  $ -2.51$     &  $-4.05$  & bridge & $-5.31$ & $-4.46$  \\
  Cl/Fe(100) & hollow &  $-2.74$   & $-3.07$    & bridge &  $-3.42$  & $-3.20$  \\
  N/Sr(111) & fcc &    $-6.65$     & $-6.48$    & hcp  &  $-6.77$   & $-6.29$   \\
  O/Sr(111) & fcc &    $-8.06$     & $-7.64$    & hcp  &  $-8.45$    & $-8.36$  \\
  F/Sr(111) & fcc &    $-6.82$     & $-6.71$    & hcp  &  $-7.01$  & $-6.78$    \\
  Cl/Sr(111) & fcc &   $-5.34$     & $-5.31$    & hcp  &  $-5.47$    & $-5.35$  \\
  F/Ru(001) & fcc &  $-4.03$     & $-3.54$ & bridge &   $-4.14$ &  $-3.59$  \\
  F/W(110)& hollow  &  $-4.77$   & $-4.60$   & top    &  $-5.15$  & $-4.87$  \\
  F/Os(001) & fcc &  $-3.66$     &  /      & top &  $-4.10$   &  $-3.98$ \\
  F/Ir(111) & fcc  &  $-3.12$     &  /     & top &  $-3.64$   &  $-3.42$ \\
  Cl/Ir(111)& fcc &  $-3.04$     &  /      & top &  $-3.05$   &  $-2.07$ \\                                                          
  F/Pt(111) & fcc  &  $-2.62$     &  /     & top &  $-3.13$   &  $-2.74$ \\
  F/Bi(100) & hollow &    /      & $-3.20$ & top    &   $-4.16$ & $-3.36$ \\
    \hline
\end{tabular}


  \label{tab:exceptions}
\end{table*}

\clearpage

